\documentclass[aps,prd,nopacs,floatfix,notitlepage,nofootinbib,twocolumn,a4paper,longbibliography]{revtex4-1}

\usepackage{amsfonts,amsmath,units,wasysym,epsfig,graphicx,verbatim,color,subfigure,graphicx,bm,mathrsfs,lipsum,hyperref,cleveref}
\usepackage{booktabs}
\usepackage[normalem]{ulem}  
\usepackage{multirow}
\begin{document}

\newcommand{\bhaskar}[1]{\textcolor{blue}{ \bf BB: #1}}
\newcommand{\SR}[1]{\textcolor{red}{ \bf SR: #1}}
\newcommand{\OKC}{The Oskar Klein Centre, Department of Astronomy, Stockholm University, AlbaNova, SE-10691 Stockholm,
Sweden}

\newcommand{\HU}{{Hamburger Sternwarte, Gojenbergsweg 112, D-21029 Hamburg, Germany}}

\title{Simultaneously Constraining the Neutron Star Equation of State and Mass Distribution through Multimessenger Observations and Nuclear Benchmarks}

\author{Bhaskar Biswas$^{\rm 1, \rm 2}$, Stephan Rosswog$^{\rm 1, \rm 2}$}
\affiliation{$^{\rm 1}$\HU, $^{\rm 2}$\OKC}

\begin{abstract}

With ongoing advancements in nuclear theory and experimentation, together with a growing body of neutron star (NS) observations, a wealth of information on the equation of state (EOS) for matter at extreme densities has become accessible. Here, we utilize a hybrid EOS formulation that combines an empirical parameterization centered around the nuclear saturation density with a generic three-segment piecewise polytrope model at higher densities.  We incorporate data derived from chiral effective field theory ($\chi$EFT), perturbative quantum chromodynamics (pQCD), and from experiments such as PREX-II and CREX. Furthermore, we examine the influence of a total of 70 NS mass measurements up to April 2023, as well as simultaneous mass and radius measurements derived from the X-ray emission from surface hot spots on NSs. Additionally, we consider constraints on tidal properties inferred from the gravitational waves emitted by coalescing NS binaries. To integrate this extensive and varied array of constraints, we utilize a hierarchical Bayesian statistical framework to simultaneously deduce the EOS and the distribution of NS masses. We find that incorporating data from $\chi$EFT significantly tightens the constraints on the EOS of NSs near or below the nuclear saturation density. However, constraints derived from pQCD computations and nuclear experiments such as PREX-II and CREX have minimal impact. Taking into account all available data, we derive estimates for key parameters characterizing the EOS of dense nuclear matter. Specifically, we determine the slope ($L$) and curvature ($K_{\rm sym}$) of the symmetry energy to be $54^{+10}_{-10}$ MeV and $-158^{+73}_{-63}$ MeV with 90\% credibility, respectively. Additionally, we infer the radius and tidal deformability of an NS with a mass of 1.4 solar masses ($M_{\odot}$) to be $12.34_{- 0.53}^{+0.43}$ km and $436_{-117}^{+109}$, respectively. Furthermore, we estimate the maximum mass of a non-rotating NS to be $2.22_{-0.19}^{+0.21} M_{\odot}$ with 90\% credibility. 
\end{abstract}

\maketitle

\section{Introduction}

The study of neutron stars (NSs) offer unique insights into the behavior of matter under extreme conditions, where the densities are greater than those found in atomic nuclei. 
To determine the internal composition and structure of a NS
one needs to know, apart from the neutron star mass, the equation of state (EOS) that governs matter at supra-nuclear densities~\cite{Lattimer_2016,Oertel_2017,Baym_2018}. The EOS, which encapsulates the relationship between pressure and density within the star, is crucial for predicting neutron star key characteristics such as radii, moments of inertia, and tidal deformabilities~\cite{Hinderer:2007mb},
as well as for the dynamics and final fate of a neutron star merger~\cite{Baiotti:2019sew, shibata19,bernuzzi20}. However, these characteristics are not solely functions of the EOS; they also depend intricately on the star's mass. Consequently, the accurate determination of NS properties necessitates a simultaneous inference of both mass and EOS.

Thanks to the LIGO/Virgo/Kagra collaboration~\citep{advanced-ligo,advanced-virgo}, we have now entered into the era of gravitational wave (GW) astronomy and, up to now, two binary neutron star (BNS) merger events named GW170817~\citep{TheLIGOScientific:2017qsa,Abbott:2018exr} and GW190425~\citep{Abbott:2020uma} have been detected. In addition to GW observations, electromagnetic (EM) observations of NSs, including radio and X-ray data, contribute significantly to our understanding. Notably, recent simultaneous measurements of mass and radius for pulsars like PSR J0030+0451~\citep{Riley:2019yda,Miller:2019cac}, PSR J0740+6620~\citep{Riley:2021pdl,Miller:2021qha}, and PSR J0437+4715~\cite{Choudhury:2024xbk} using the Neutron Star Interior Composition Explorer (NICER)~\citep{2016SPIE.9905E..1HG}  instrument have imposed substantial constraints on the high-density portion of the NS EOS. Finally with advancements in pulsar radio timing and X-ray observations, the increasing number of NSs with accurately measured masses over the past few decades enables the statistical inference of mass distribution features.

   Over the recent decades, ab-initio many-body approaches, like the chiral effective field theory ($\chi$EFT)~\cite[see, e.g.,][]{Epelbaum:2008ga,Machleidt:2011zz,Hammer:2012id,Hebeler:2020ocj}, have gained significant traction in quantifying theoretical uncertainties concerning the nuclear EOS. While sophisticated calculations exist in the literature for finite nuclei and nuclear matter, the estimation breaks down beyond $2 n_0$~\cite{Drischler:2021kxf}, where $n_0= 0.16$ fm$^{-3}$ is the nuclear saturation density. But it provides an excellent check on the EOSs inferred from the NS observations. While we try to infer the high-density EOS directly from NS observations, a complementary description of NS matter comes from perturbative quantum chromodynamics (pQCD) calculations which become reliable at sufficiently high densities beyond $\sim 40 n_0$~\cite{Komoltsev:2021jzg}. Terrestrial experiments also provide us with important information about subnuclear matter below $n_0$. For example, the latest neutron skin thickness measurements of $\rm{Pb}^{208}$~\cite{PREX:2021umo}, $\rm{Ca}^{40}$~\cite{CREX:2022kgg} nuclei have provided information about the nuclear symmetry energy and its slope parameter.

   The  datasets that we employ here provide information on the EOS at different densities~\cite{Raaijmakers:2019dks,Landry_2020PhRvD.101l3007L,Jiang:2019rcw,Traversi:2020aaa,Biswas:2020puz,Biswas:2020xna, Dietrich:2020efo,Miller:2021qha,Biswas:2021paf,Biswas:2021pvm, Biswas:2021yge,Ghosh:2021bvw,Tiwari:2023tkj}. The heaviest pulsar mass measurements help to eliminate EOSs that cannot support the observed NS masses; these constraints tend to most significantly impact inference near $4-6\,n_0$ and they typically favour a stiffer EOS. The X-ray and GW observations provide constraints on the NS mass, radius, and tidal deformabilities; and give information about the EOS mainly in the region $1-4\,n_0$. In contrast, nuclear experiments or calculations are primarily influential in constraining the EOS near or below $n_0$. Finally, pQCD calculations will constrain the EOS starting from only a few times the nuclear saturation density  $ > 2.2\,n_0$~\cite{Komoltsev:2021jzg}. 

   The need for simultaneous mass and EOS inference under a Bayesian framework arises from the need to connect observational data and theoretical models. Bayesian inference provides a systematic method to update our prior knowledge based on new data, accommodating uncertainties and correlations inherently present in the data. If we were to infer mass and EOS independently, we might introduce biases because the parameters are not independent; they are degenerate. This means that different combinations of mass and EOS can produce similar observational signatures. Ignoring this degeneracy leads to incorrect or biased posterior distributions. By simultaneously inferring mass and EOS, we construct a joint posterior distribution that correctly accounts for the uncertainties and correlations between these parameters. This approach ensures that the derived properties are consistent with both the observed data and the physical models. Simultaneous inference allows for more accurate predictions of NS properties. For instance, predicting the maximum mass or radius of NSs relies on an accurate understanding of the EOS, which in turn depends on correctly modeling the mass distribution.

    This article is organized as follows: In Sec~\ref{section: EOS}, we provide a description of the construction of the EOS parameterization. We discuss the hybrid parameterization approach that combines nuclear empirical parameters at low densities with piecewise polytropic (PP) models at higher densities. Sec~\ref{section: Methodology} delves into the multi-messenger observations and the theoretical models employed to constrain the EOS. This section also outlines the hierarchical Bayesian framework used to integrate various datasets. In Sec~\ref{sec:GMM}, we employ Gaussian Mixture Models to robustly analyze multi-dimensional observational data from gravitational wave detections, X-ray observations, and empirical EOS parameters, optimizing model selection to capture the complex dependencies within these datasets. Sec~\ref{section: results} presents the results obtained from the combined analysis of theoretical predictions, observational data, and experimental constraints. We detail the inferred properties of NSs, including the symmetry energy parameters, radius, and tidal deformability. Sec~\ref{sec:systematics} addresses the various systematic uncertainties that can affect the inference of the equation of state (EOS), highlighting their impact on the accuracy and reliability of EOS determinations. Finally, Sec~\ref{section: conclusion} summarizes our findings and discusses their implications for future research, highlighting the synergy between different types of data in enhancing our understanding of NS interiors.
   
\section{EOS and mass distribution model}
\label{section: EOS}
To approach the EOS of dense matter, one must aim to remove the systematic uncertainties of the modelling as much as possible to interpret the data properly. The central theme of this paper is to construct a consistent framework to connect the observational results from astrophysical detections to microscopic properties of dense matter combined with theoretical prediction of nuclear matter at low-density and high-density coming from $\chi$EFT and pQCD calculations, respectively. To interpret all these different types of datasets, we use a nuclear-physics motivated hybrid parameterization for NS matter which is a combination of two widely used models briefly discussed below in subsection~\ref{subsection:empirical-EOS} and~\ref{subsection:PP-EOS}. This hybrid EOS parameterization was developed by \textcite{Biswas:2020puz} and has been utilized in multi-messenger analyses of NS properties \cite{Biswas:2020xna,Biswas:2021yge}. Additionally, it has been applied to create a framework for inferring the Hubble parameter directly from GW signals from future BNS mergers \cite{Ghosh:2022muc,Ghosh:2024cwc}.  We review the construction and necessary modification of this hybrid parameterization for NS matter in subsection~\ref{subsection:nuclear+PP-EOS} and the mass distribution model in subsection~\ref{subsection: nspop}. 

\subsection{Nuclear empirical parameterization:} 
\label{subsection:empirical-EOS}
The EOS of nuclear matter around $n_0$ is described in terms of the nuclear empirical parameters~\cite{PhysRevC.44.1892,Haensel:2007yy,Baldo:2016jhp}. These are defined from the parabolic expansion of the energy per nucleon  $e(n,\delta)$ of asymmetric nuclear matter as:
\begin{equation}
    e(n,\delta) \approx  e_0(n) +  e_{\rm sym}(n)\delta^2,
\end{equation}
where \(n\) is the baryon number density, $e_0(n)$ is the energy per nucleon in symmetric nuclear matter
that contains equal numbers of neutrons and protons,
$e_{\rm sym}(n)$ is the nuclear symmetry energy, the energy cost for
having asymmetry in the number of neutrons and protons in the system
and $\delta=(n_n-n_p)/n$ is the measure of this asymmetry, with \(n_n\) and \(n_p\) being the neutron and proton densities, respectively. Mathematically, symmetry energy can be expressed as the difference between the nuclear energy per particle in pure neutron matter (PNM) and symmetric nuclear matter (SNM),
\begin{equation}
    e_{\rm sym}(n) = e_{\rm PNM}(n)-e_0(n)\,.
    \label{eq:S0}
\end{equation}
One can expand $e_0(n)$ and $e_{\rm sym}(n)$ around the
saturation density $n_0$ as:
\begin{eqnarray}
 e_0(n) &=&  e_0(n_0) + \frac{ K_0}{2}\chi^2 \label{eq:e0} +\,...,\\
e_{\rm sym}(n) &=&  e_{\rm sym}(n_0) + L\chi + \frac{ K_{\rm sym}}{2}\chi^2 
+ ..., \label{eq:esym}
\end{eqnarray}  
where:
\begin{itemize}
    \item $\chi \equiv (n-n_0)/3n_0$ quantifies the deviation from saturation density.
    \item \(e_0(n_0)\) is the energy per nucleon at saturation density.
    \item \(K_0\) is the nuclear incompressibility modulus, describing the curvature of the energy per nucleon at \(n_0\).
    \item $e_{\rm sym}(n_0)$ is the symmetry energy at saturation density.
    \item \(L\) is the slope parameter of the symmetry energy at \(n_0\).
    \item \(K_{\text{sym}}\) is the curvature parameter of the symmetry energy at \(n_0\).
\end{itemize}

The pressure \(p\) and energy density \(\epsilon\) are derived from the energy per nucleon and its derivatives\footnote{Similarly, the individual pressures and energy densities of PNM and SNM can also be calculated.}. The total energy density \(\epsilon\) includes the rest mass energy. They are given by:

\begin{eqnarray}
\epsilon = n(e+m_Nc^2) \,,
\nonumber \\
p = n^2\frac{\partial e}{\partial n}
 \,.
\label{eq:eostotal}
\end{eqnarray}
where \(m_N\) and $c$ are the average nucleon mass and speed of light, respectively. We neglect the contribution of muons, as their effect on the extracted symmetry energy
and its derivatives is minimal~\cite{Essick:2021kjb}. Therefore, due to charge neutrality, the electron density in $\beta$-equilibrium is equal to the proton density, $n_{\rm e} = Y_e^{\rm eq}(n) \cdot n$.
 We evaluate the electron fraction by implementing $\beta$ equilibrium (for a derivation of these relations, see, e.g.,~\cite{Blaschke:2016lyx,Raithel:2019gws}):
\begin{equation}
\label{eq:YpBInv}
Y_e^{\rm eq} = \frac{1}{2} + \frac{(2 \pi^2)^{1/3}}{32} \frac{n}{\xi} \left\{ (2\pi^2)^{1/3} - \frac{\xi^2}{n} \left[\frac{\hbar c}{e_{\rm sym}(n)}\right]^3 \right\},
\end{equation}
where, for simplicity, we have introduced the auxiliary quantity $\xi$, defined as
\begin{multline}
\label{eq:xi}
\xi \equiv \left[ \frac{e_{\rm sym}(n)}{\hbar c} \right]^2  \times \\
\left\{ 24 n \left[ 1+ \sqrt{ 1 +  \frac{\pi^2 n}{288}\left(\frac{\hbar c}{e_{\rm sym}(n) }\right)^3}\right]  \right\}^{1/3}.
\end{multline}  
$\hbar$ is the Planck constant divided by $2\pi$.

\subsection{Piecewise Polytropic parameterization:}
\label{subsection:PP-EOS}
In the piecewise polytropic (PP) model~\cite{Read:2008iy}, the EOS is divided into several segments, each described by its own polytropic relation. Compare to a single polytrope, this allows for a more flexible and accurate representation of the NS EOS across different density ranges. At baryon number densities $n > n_1$ we use a set of three piecewise polytropes
$P_i(n)$~\cite{Hebeler:2013nza} to describe homogeneous matter in
beta-equilibrium:
\begin{subequations}
  \begin{gather}
    \nonumber\\[-0.5\baselineskip]
    p(n) = \begin{cases}
      p_1(n) & n_1 < n \leq n_2,\\
      p_2(n) & n_2 < n \leq n_3,\\
      p_3(n) & n_3 < n,
    \end{cases}\\
    \begin{aligned}
      p_i(n) &= K_i n^{\Gamma_i}, &
      K_i &= \frac{P(n_i)}{n_i^{\Gamma_i}}.
    \end{aligned}
  \end{gather}
\end{subequations}
This introduces six new parameters: $\{n_1, n_2, n_3, \Gamma_1, \Gamma_2, \Gamma_3\}$, with the coefficients $K_i$ fixed by demanding that $P(n)$ be continuous.

\subsection{Hybrid nuclear+PP parameterization:} 
\label{subsection:nuclear+PP-EOS}

Our hybrid EOS parameterization is divided into three parts:

\textbf{Density below $n_{\rm match}$:} 
As the choice of the crustal EOS does not significantly influence the NS observable \cite{Biswas:2019ifs,Gamba:2019kwu}, we employ a fixed SLy crust model~\cite{Douchin:2001sv} below $n_{\rm match}$, where $n_{\rm match}$ represents the junction density above which the empirical parameterization is utilized. The determination of $n_{\rm match}$ ensures the continuity of pressure and energy density between the SLy EOS and the empirically parameterized EOS. The crust model is parameterized using a PP approach. The optimal parameters for the analytical representation of the SLy EOS are taken from Ref.~\cite{Read:2008iy}.

\textbf{Density between $n_{\rm match}$ to $1n_0 \leq n_1 \leq 2n_0 $:} Chiral effective field theory constraints are reliable up to approximately twice the nuclear saturation density. However, the precise breakdown scale of $\chi$EFT remains uncertain. Recently, Tiwari et al.~\cite{Tiwari:2023tkj} conducted a Bayesian analysis that combined astrophysical observations with $\chi$EFT constraints, employing two different transition densities, \(1.5n_0\) and \(2n_0\). Their study revealed that adopting \(2n_0\) as the breakdown density yields tighter posteriors. Given the current arbitrariness in selecting the breakdown density, we marginalize over it within the range from \(n_0\) to \(2n_0\). 
To incorporate this, we introduce a free parameter \(n_1\) into our model which represents the $\chi$EFT breakdown density. In the density range below \(n_1\), we use a nuclear empirical parameterization described in Sec.~\ref{subsection:empirical-EOS}. The uncertainties in the lower-order empirical parameters are small and well-constrained~\cite{Brown:2013pwa,Tsang:2019ymt}. Consequently, we fix the lowest-order parameters, such as the saturation density $n_0= 0.16$ fm$^{-3}$ and the energy per nucleon at saturation density, \(e_0(n_0) = -15.9~\mathrm{MeV}\). A survey of 53 experimental results in 2016~\cite{Oertel:2016bki} found symmetry energy and compressibility values of \( e_{\rm sym} (\rho_0) = 31.7 \pm 3.2 \) MeV and \( K_0 = 240 \pm 30 \) MeV. While there are larger uncertainties on these parameters, previous subsequent Bayesian analyses~\cite{Biswas:2020puz,Biswas:2021yge,Char:2024kgo}, which combined multiple astrophysical observations (gravitational waves and X-ray data), found that current observational constraints do not provide significant information on these parameters. Given this, we chose to fix \( e_{\rm sym} (\rho_0) \) and \( K_0 \) at their median values from the survey---31.7~MeV and 240~MeV, respectively---rather than unnecessarily increasing the dimensionality of the EOS parameterization. Since the present data does not favor or constrain a particular range of \( e_{\rm sym} \) and \( K_0 \), our results remain robust under reasonable variations of these parameters. This leaves the slope parameter \(L\) and the curvature parameter \(K_{\rm sym}\) of the symmetry energy as the two remaining free parameters in the model.

\textbf{Density above $n_1$:} For higher densities the EOS is extended by using a PP EOS with three segments. The parameter ranges for the polytropic exponents are $1.0 \le \Gamma_{1} \le 6$, $0 \le \Gamma_{2} \le 8.5$, and $0.5 \le \Gamma_{3} \le 8.5$. The densities between the polytropes are $n_1 \le n_{2} < n_{3} < n_{\text{max}}$, where $n_{\text{max}} \approx 8.3 \, n_{0}$ is chosen to be the maximum central density reached. 

In total there are 8 free parameters ($\theta_\epsilon$) in hybrid+PP EOS parameterization: 2 from nuclear empirical parameterization ($L$ and $K_{\rm sym}$) and 6 from PP parameterization ($n_{1}$, $n_{2}$, $n_{3}$, $\Gamma_{1}$, $\Gamma_{2}$, and $\Gamma_{3}$). For each EOS parameters, we assume a uniform prior and their ranges are shown in Table~\ref{tab:prior}. We have taken broader ranges for all the parameters to ensure flexibility in the equation of state and to avoid scenarios where priors could strongly constrain the posteriors. This approach allows the data to guide the inference without imposing unnecessary restrictions. Similar strategies have been adopted in previous studies, such as Refs.~\cite{Hebeler:2013nza,Raaijmakers:2019dks}.
 Finally, it is crucial to ensure that the parameterized EOS complies with fundamental physical principles, specifically causality and the monotonicity of pressure with respect to density. The speed of sound within the neutron star (NS) must not exceed the speed of light to adhere to the principles of special relativity.  Mathematically, this condition is expressed as:
\begin{equation}
\left(\frac{\partial p}{\partial \epsilon}\right)_{S} \leq c^2,
\end{equation}
where the derivative is computed at constant entropy $S$. The pressure must monotonically increase with density to ensure hydrostatic stability within the NS (``Le Chatelier's principle", see e.g. \cite{Glendenning:1997wn}):

\begin{equation}
\left(\frac{\partial p}{\partial n}\right)_{S} > 0.
\end{equation}

\subsection{Mass distribution model}\label{subsection: nspop}

Previous studies~\cite{Ozel:2012ax, Antoniadis:2016hxz, Alsing:2017bbc,Farrow:2019xnc,Farr2020,Shao:2020bzt,Das:2024ams} indicate that the galactic population of NS masses follow a double Gaussian distribution. However there is no {\it a priori} reason to believe that it would be same in other galaxies too. NSs detected through GWs may follow a different distribution than the galactic mass distribution~\cite{Landry:2021hvl}. Nevertheless, given the small number of NSs detected through GWs, a wrong choice of population will have negligible effect on the EOS inference. Therefore we assume for simplicity that all NSs observed in the universe follow a bimodal mass distribution:

\begin{multline}
        P_\textsc{nn}(M|\mu_1,\sigma_1,\mu_2,\sigma_2,w,M_{\rm min},M_{\rm max}) = \\
        \left[w \mathcal{N}(M|\mu_1,\sigma_1)/B 
        + (1-w) \mathcal{N}(M|\mu_2,\sigma_2)/C\right] \\
        U(M|M_{\rm min},M_{\rm max}) , \label{bimodal}
\end{multline}
We fix $M_{\rm min}=0.9\,M_{\odot}$ to reduce the dimensionality of the parameter space. We do not consider the possibility of low-mass NSs, which is theoretically disfavoured by the early evolutionary stages of a NS~\cite{Strobel:1999vn} and is consistent with the minimum remnant masses predicted by core-collapse supernova simulations ~\cite{Janka:2007di, Fischer_2010, Radice:2017ykv, Suwa:2018uni}. $M_\mathrm{max}$ is the maximum mass for that particular EOS and the term $U(M|M_{\rm min},M_{\rm max})$ reads:
\begin{equation}
    U(M|M_{\rm min},M_{\rm max}) = 
    \begin{cases} 
        \frac{1}{M_\mathrm{max} - M_\mathrm{min}} & \text{if } M_\mathrm{min} \leq M \leq M_\mathrm{max}, \\
        0 & \text{else.}
    \end{cases}
\end{equation}
Here the normalization factor is important as it prefers the EOS with slightly larger $M_\mathrm{max}$ than the heaviest observed NS mass and disfavor EOS with much larger $M_\mathrm{max}$. The ranges of population model parameters ($\theta_m$, see Eq.~\ref{bayes theorem}) are listed in Table~\ref{tab:prior}.

\begin{table}[ht!]
\begin{center}
\caption{Prior ranges of the NS EOS and mass distribution model hyperparameters used in this work. The notation $U$ stands for uniform distribution.}
\label{tab:prior}
\begin{tabular}{ |c|c|c|c|c| } 
  \hline
  Model & Parameters & Units & Prior \\  
  \hline\hline
  \multirow{7}{3.2em}{EOS}  
                         & $L$ & MeV &   $U(0,150)$\\
                         & $K_{\rm sym}$ & MeV &  $U(-600,100)$\\
                         & $n_{1}$ & $n_0$  & $U(1,2)$\\ 
                         & $n_2$ & $n_0$   & $U(n_1 ,5)$\\
                         & $n_3$ & $n_0$   & $U(n_2 ,8)$\\
                         & $\Gamma_{1}$ &  -  & $U(1,6)$\\
                         & $\Gamma_{2}$ &  -  & $U(0,8.5)$\\
                         & $\Gamma_{3}$ &  -  & $U(0.5,8.5)$\\
                         
  \hline
  \multirow{4}{3em}{Mass} & $\mu_1$ & $M_{\odot}$ &  $U(0.9,\mu_2)$\\ 
                          & $\sigma_1$ & $M_{\odot}$  &   $U(0.01,\sigma_2)$\\
                          & $\mu_2$ & $M_{\odot}$ &  $U(0.9,M_{\text{max}})$\\ 
                          & $\sigma_2$ & $M_{\odot}$  &   $U(0.01,1.0)$\\
                          & $w$ & -  &   $U(0.1,0.9)$\\
  \hline
\end{tabular}
\end{center}
\end{table}

\section{Bayesian methodology}
\label{section: Methodology}

The joint posterior distribution of the EOS and population hyperparameters (denoted as $\theta$) are computed using nested sampling algorithm implemented in~{\tt Pymultinest}~\citep{Buchner:2014nha}:
\begin{equation}
    P(\theta | {D}) = \frac{P ({D} | \theta) \times P(\theta)}{P(D)}\, = \frac{\Pi_i P ({D_i} | \theta) \times P(\theta)}{P(D)}\,,
    \label{bayes theorem}
\end{equation}
where $\theta = ( \theta_\epsilon, \theta_m)$ is a set of hyper-parameters corresponds to different parameterized models that describe the EOS and mass distribution of NSs respectively, $D$ is the set of data from nuclear theory, experiments, and the different types of astrophysical observations that are used to construct the likelihood, $P(\theta)$ are the priors of those parameters and $P(D)$ is the Bayesian evidence, given the particular EOS model. The expressions of individual likelihood are described below. 

\begin{figure*}[ht!]
    \centering
    \begin{tabular}{cc}
    \includegraphics[height=0.3\textheight,width=0.45\textwidth]{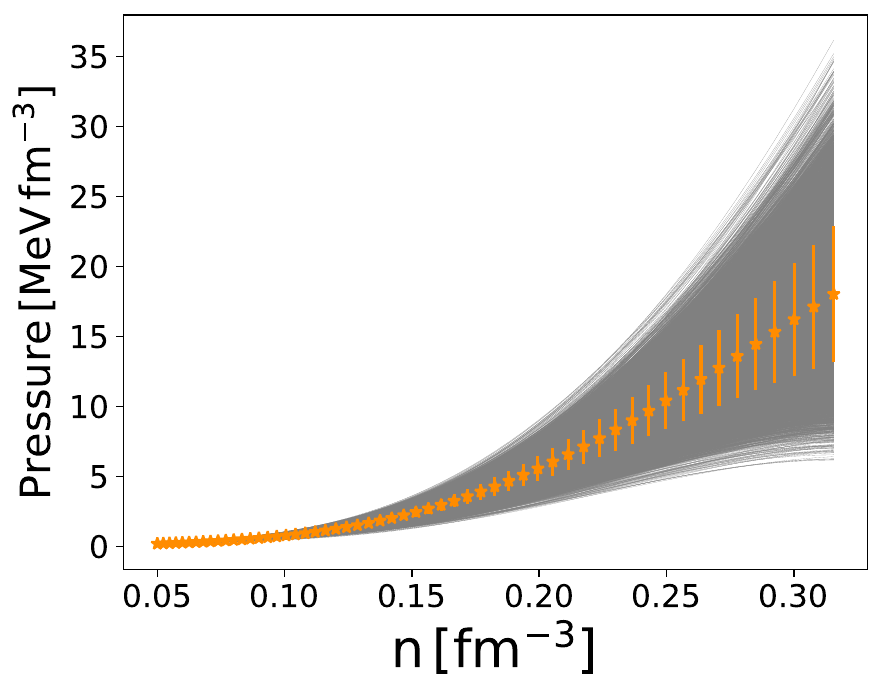} &
    \includegraphics[height=0.3\textheight,width=0.45\textwidth]{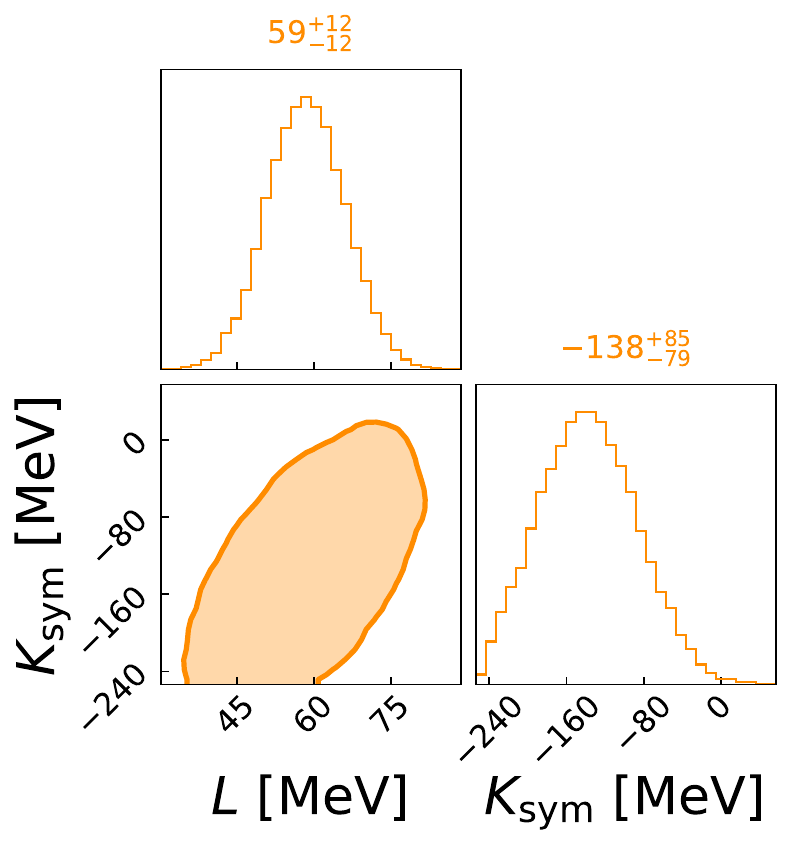}
    \\
    \end{tabular}
\caption{In the left panel, the mean values and standard deviation of PNM pressure are shown within a density range between $0.5 n_0$ and $2 n_0$ using orange colour. Corresponding 20000 random draws are shown in grey lines. The fitted 1D distributions of the slope ($L$) and curvature ($K_{\rm sym}$) of the symmetry energy, and their correlation is shown at the right panel. In the marginalized one-dimensional plot, the median and 90\% CI values of $L$ and $K_{\rm sym}$ are shown.}
    \label{fig:chiral-post}
\end{figure*}
\subsection{$\chi$EFT Likelihood}

In the relevant region, typically below about $2n_0$, $\chi$EFT, utilizing pions and nucleons as degrees of freedom \cite{Epelbaum:2008ga, Machleidt:2011zz, Hammer:2019poc, Tews:2020hgp}, has become the leading microscopic approach for understanding nuclear interactions. This methodology has played a crucial role in advancing our understanding of the EOS for infinite nuclear matter and the internal composition of NSs. Notably, it provides predictions with quantifiable theoretical uncertainties (for recent reviews, see Refs.~\cite{Hebeler:2015hla, Drischler:2019xuo, 10.3389/fphy.2019.00213, Drischler:2021kxf}).

A significant stride in comprehending the uncertainties in EOS was recently made by the \emph{Bayesian Uncertainty Quantification: Errors in Your EFT} (BUQEYE) collaboration \cite{BUQEYEgithub}. This collaboration introduced a Bayesian framework \cite{Drischler:2020hwi, Drischler:2020yad} for quantifying and addressing correlated errors arising from the truncation of effective field theory (EFT) in calculations of infinite nuclear matter, utilizing Gaussian Processes. They performed a statistical analysis of the zero-temperature EOS based on nucleon-nucleon and three-nucleon interactions within $\chi$EFT. The results included inferred posterior distributions for nuclear saturation properties and essential quantities relevant to NSs, such as the nuclear symmetry energy and its density dependence.

This initiative was instigated by recent advancements in many-body perturbation theory~\cite{Drischler:2017wtt}, which facilitated improved $\chi$EFT predictions for the pure neutron matter EOS and first-order-by-order calculations in symmetric nuclear matter up to next-to-next-to-next-to-leading order (N3LO) in the chiral expansion \cite{Drischler:2017wtt, Leonhardt:2019fua, Drischler:2020hwi}.

Using the Jupyter notebooks~\cite{BUQEYEgithub} provided by the BUQEYE collaboration, we extract the mean values, standard deviations (encoding the EFT truncation errors), and correlation information of the energy per particle, pressure, and speed of sound in PNM and SNM, and also the symmetry energy. We draw a sufficiently large number (20,000) of PNM pressure versus density curves \((\mathbf{p}) = (p_{1:n})\) between \(0.5 n_0\) and \(2 n_0\) from an \(n\)-dimensional multivariate normal distribution with mean \((\bar{\mathbf{p}}) = (\bar{p}_{1:n})\) and covariance matrix \(\mathbf{C} = C_{1:n,1:n}\). This sample size ensures robustness and accuracy in our analysis without affecting the results of the calculations. A simple way of generating a random draw $\mathbf{x}$ from a multivariate normal distribution of mean $\boldsymbol{\mu}$ and covariance $\mathbf{C}$ is to generate a random vector $\mathbf{n}$ with elements drawn from a unit normal and then ii) set $\mathbf{x} = \boldsymbol{\mu} + \mathbf{L} \, \mathbf{n}$, where $\mathbf{L}$ is the Cholesky decomposition of $\mathbf{C}$ (i.e., a lower triangular matrix such that $\mathbf{L} \, \mathbf{L}^{\rm T} = \mathbf{C}$). In the left panel of Fig.~\ref{fig:chiral-post}, the mean values and standard deviation of PNM pressure and its corresponding 20,000 random draws are shown. We fit each of these PNM pressure versus density curves with the nuclear empirical paremeterization described in Section~\ref{subsection:empirical-EOS}. The fitted 1D distributions of empirical EOS parameters $L$ and $K_{\rm sym}$, and their correlation is shown in  the right panel of Fig.~\ref{fig:chiral-post}. 

\subsection{NS mass Likelihood}
\label{subsection: mass-like}

\citet{Alsing:2017bbc} use three classes of pulsar mass measurements. The first consists of well-constrained pulsar mass measurements, where the likelihood function is assumed to be (proportional to) a Gaussian:
\begin{equation}
    P\left( D_p \mid M_p \right) \propto \exp\left( - \frac{\left( M_p - \mu \right)^2}{2 \sigma^2} \right),
\end{equation}
with \(D_p\) the (abstract) pulsar measurement data, and \(M_p\) the true mass of the pulsar. Here \(\mu\) is the reported mass measurement and \(\sigma\) is its uncertainty.

The second class consists of measurements of the mass function,
\begin{equation}
    f \equiv \frac{M_p q^3 \sin^2 \iota}{\left( 1 + q\right)^2},
\end{equation}
where \(q = M_c / M_p\) (\(M_c\) is the companion mass; this implies \(0 < q < \infty\)) and \(\iota\) is the angle between the orbital angular momentum and the line of sight. \(f\) is assumed to be measured perfectly (uncertainties on \(f\) are below the part-per-thousand level, and therefore ignorable), and measurements of the total mass of the binary system, \(M_t = M_p \left( 1 + q \right)\), with an assumed Gaussian likelihood centered at the measured total mass, \(\mu_t\), with standard deviation \(\sigma_t\). The complete likelihood is
\begin{multline}
    P\left( D_p \mid M_p, M_t, \iota \right) \propto \delta\left( f\left( M_p, M_t, \iota \right) - f_p \right) \times \\
    \exp\left( -\frac{\left( M_t - \mu_t\right)^2}{2 \sigma_t^2}\right),
\end{multline}
with \(f_p\) the measured mass function. Integrating over \(\iota\) with an isotropic prior (flat in \(\cos \iota\)) leaves the marginal likelihood
\begin{multline}
    \label{eq:marginal-like-mt}
    P\left( D_p \mid M_p, M_t \right) \propto \exp\left( - \frac{\left( M_t - \mu_t \right)^2}{2 \sigma_t^2} \right) \times \\ \frac{M_t^{4/3}}{3 \left( M_t - M_p \right)^2 f_p^{1/3} \sqrt{1 - \frac{f_p^{2/3} M_t^{4/3}}{\left(M_t - M_p\right)^2}}}.
\end{multline}

The third class consists of systems with (perfect) measurements of the mass function and Gaussian uncertainties on the mass ratio, \(q\); similarly, integrating over \(\iota\) with an isotropic prior leads to a marginal likelihood
\begin{multline}
    \label{eq:marginal-like-q}
    P\left( D_p \mid M_p, q \right) \propto  \exp\left( - \frac{\left( q - \mu_q \right)^2}{2 \sigma_q^2} \right) \times \\
    \frac{\left( 1 + q \right)^{4/3}}{3 f_p^{1/3} M_p^{2/3} q^2 \sqrt{1 - \left(\frac{f_p}{M_p} \right)^{2/3} \frac{\left(1 + q\right)^{4/3}}{q^2}}};
\end{multline}
this is Eq.\ \eqref{eq:marginal-like-mt} with \(M_t \to M_p \left( 1 + q \right)\) in the term that arises from integrating over \(\iota\). Notably, we use the expression from Ref.~\cite{Farr2020}, which corrects the typos present in \citet{Alsing:2017bbc} Eq.\ (4).

We take the measurements of the NS masses (till April 2023) from Ref.~\cite{Fan:2023spm}, a total of 129 items~\footnote{Note that Ref.~\cite{Fan:2023spm} has a list of 136 NS mass measurements. But we excluded the mass measurements of NSs coming from two GW and three NICER observations to avoid double-counting.} classified into three types. Among them, 106 items are the individual estimates of masses of NSs, while 17 of them are the measurements of mass function and total mass of binary systems. The other 7 items are the measurements of mass function and mass ratio of the binary systems. To approximate well the non-Gaussian mass measurements in Table I of Ref.~\cite{Fan:2023spm}, we use the asymmetric normal distribution studied in Refs.\citep{Kiziltan:2013oja} to reproduce the error distribution, of which the density function is given by
\begin{multline}\label{skew-normal-errors}
    {\rm AN}(w\mid c,d) = \frac{2}{d(c+\frac{1}{c})} \left\{ \phi\left(\frac{w}{cd}\right) 1_{[0,\infty)}(w) \right. \\
    \left. + \phi\left(\frac{c w}{d}\right) 1_{(-\infty,0)}(w) \right\},
\end{multline}
where $c > 0$, $d > 0$, $\phi$ means normal distribution, and $1_{A}(\cdot)$ denotes the indicator function of set $A$, which equals 1 if the argument is in $A$ and 0 otherwise. Thus given the NS mass measurements ${\mathcal{M}_{i}}_{-\ell_{i}}^{+u_{i}}$ ($+u_{i}/-\ell_{i}$ are 68\% central limits), parameters $c_{i}$ and $d_{i}$ for the $i$th NS can be estimated through $c_{i}=$ $(u_{i}/\ell_{i})^{1/2}$ and $\int_{-\ell_{i}}^{u_{i}} \text{AN}(w\mid c_{i},d_{i}) \text{d}w = 0.68$. Then it is straightforward to calculate the probability for a specific pulsar mass $M_{\rm p}$ via $P(D^i\mid M_{\rm p})=\text{AN}(M_{\rm p}-\mathcal{M}_{i}\mid c_{i}, d_{i})$. For the data only having mass function $f$ and mass ratio $q$ (or total mass $M_{\rm T}$) measurements available in Table I of Ref.~\cite{Fan:2023spm}, we adopt the Eqs.~\ref{eq:marginal-like-mt} and ~\ref{eq:marginal-like-q} to evaluate the probability $P(D^i\mid M_{\rm p})$. Therefore, the likelihood constructed for our inference is given by
\begin{equation}
\label{eq:likelihood}
L(D\mid \vec{\theta}) \propto \prod_{i=1}^N \int P(M_\mathrm{p} \mid \vec{\theta}) P(D^i \mid M_{\rm p}) {\rm d} M_\mathrm{p}.
\end{equation}

We classify the neutron star mass measurements from Ref.~\cite{Fan:2023spm} into highly reliable and less reliable groups. NSs in binary system with either another NS or white dwarf (WD), tend to have well-constrained masses from radio timing, especially those systems that have undergone detailed timing studies over long periods. Amongst those B1516+02B, B1802-07, and B1855+09 (NS-WD systems) are moderately reliable, their measurements come with larger uncertainties due to the nature of their companions and system complexities. Both PSR J1811-2405 and PSR J1017-7156 have large uncertainties in their mass measurements. The broad ranges suggest challenges in constraining the system parameters accurately, making these measurements less definitive compared to systems with tighter mass constraints. Here we take a conservative approach and place these three PSRs in the less reliable category. So in total, we have 70 NS mass measurements which are highly reliable. On the other hand, black widow and redback pulsar’s mass measurements  are often less reliable due to the complexities of the companion star and the influence of irradiation, which can affect the timing and lead to larger uncertainties. Mass estimates for neutron stars in X-ray binaries can be less reliable due to difficulties in precisely measuring orbital parameters from X-ray data alone, leading to larger uncertainties. Throughout this paper, we primarily present our results based on the highly reliable PSR mass measurements. However, in Section~\ref{subsection:PSR_systematics}, we will compare the results obtained using only highly reliable PSR masses with those that include the less reliable measurements to assess the impact of these systematics on our analysis.

\subsection{GW Likelihood}

In the final inspiral stages of merging NSs, tidal interactions arising from the stars' finite size influence the binary system's orbital evolution. Each NS imparts a tidal field on its companion, which induces a quadrupole deformation and leads to an accelerated inspiral phase. The magnitude of the induced quadrupole moment is determined by the NSs' internal structure, with larger stars being more easily deformed. This effect is quantified by the dimensionless tidal deformability $\Lambda_i$ for each star $i \in \{1,2\}$, representing the ratio of the induced quadrupole moment to the external tidal field \cite{Hinderer:2007mb}. The tidal deformabilities affect the phase evolution of binaries and can therefore be constrained directly from the GW signal analysis,\cite{Flanagan:2007ix,Hinderer:2007mb}.

The advanced LIGO and Virgo detectors have detected GWs from two BNS events named GW170817 and GW190425. Each event's analysis resulted in a posterior distribution for binary parameters, including component masses $M_1$ and $M_2$ and tidal deformabilities $\Lambda_1$ and $\Lambda_2$. Table~\ref{tab:BNSdata} summarizes key properties of these events. We quote the chirp mass ${\cal{M}}$, mass ratio $q$, and the tidal parameter $\tilde{\Lambda}$ \cite{Favata:2013rwa}, which is a mass-weighted combination of $\Lambda_1$ and $\Lambda_2$. It is the best-measured tidal parameter for these events and the only one expected to be measurable with current detector sensitivities \cite{Wade:2014vqa}. In our analysis, we utilize publicly available posterior samples from \cite{170817samples} and \cite{190425samples}. The posteriors are reported with respect to a prior uniform in $\Lambda_1$, $\Lambda_2$, and the component masses in the detector frame, with the condition $M_2 \leq M_1$ \cite{Abbott:2018wiz,Abbott:2020uma}. Mass and tidal deformability measurements from both NSs in a BNS merger serve as macroscopic inputs in our inference of the EOS. The GW likelihood is expressed as:

\begin{align}
    P(D_{\mathrm{GW}}|\theta) = \int^{M_{\mathrm{max} }}_{M_2}dM_1 \int^{M_1}_{M_{\mathrm{min} }} dM_2 P(M_1,M_2|\theta)   \nonumber \\
    \times P(D_{\mathrm{GW}} | M_1, M_2, \Lambda_1 (M_1,\theta), \Lambda_2 (M_2,\theta)) \,,
    \label{eq:GW-evidence}
\end{align}
where $P(M_1,M_2|\theta)$ is the prior distribution over the component masses, which should be informed by the NS population model. Assuming a random pairing between masses in binary systems we get, 

\begin{equation}
    P (M_{1}, M_{2} \mid \theta) \propto P(M_{1} \mid \theta) P(M_{2} \mid \theta) \Theta(M_{1} > M_{2}) .
\end{equation}

Furthermore, we exclude light-curve models of electromagnetic counterparts associated with GW events due to the significant systematic uncertainties involved in interpreting kilonova physics and its relationship to the EOS (for detailed discussions, see Refs.~\cite{LIGOScientific:2017pwl, Coughlin:2018miv, Dietrich:2020efo,  Raaijmakers:2021uju,Rosswog:2022tus,Sarin:2024tja,Brethauer:2024zxg}).

\begin{table}[]
    \begin{center}
        \begin{tabular}{l @{\quad}c @{\quad}c @{\quad}c@{\quad}}
            \hline \hline
            &&&\\[-2ex]
             BNS & ${\cal{M}}$ $[M_{\odot}]$&q & $\tilde{\Lambda}$   \\[0.5ex]
              \hline 
            &&&\\[-2ex]
             GW170817~\cite{TheLIGOScientific:2017qsa,Abbott:2018wiz}& $1.186^{+0.001}_{-0.001}$ & $(0.73,1.00)$ & $300^{+500}_{-190}$   \\[0.5ex]
             GW190425~\cite{Abbott:2020uma} & $1.44^{+0.02}_{-0.02}$ & $(0.8,1.0)$ & $\lesssim 600$     \\[0.5ex]
            \hline \hline 
        \end{tabular}
    \end{center}
    \caption{
        We present a summary of the data from BNS systems used in this paper. For each parameter—chirp mass ${\cal{M}}$, mass ratio $q$, and tidal parameter $\tilde{\Lambda}$—we provide the median value (if applicable) along with the corresponding uncertainties, which are estimated at the $90\%$ credible interval (CI).
    }
    \label{tab:BNSdata}
\end{table} 

\subsection{NICER Likelihood}
\label{subsec: NICER_Likelihood}
The present study considers the mass-radius measurements of PSR J0030+0451, PSR J0740+6620, and PSR J0437+4715, as reported by the NICER collaboration. Initially, the Bayesian inference of the radius and mass of PSR J0030+0451 was independently performed by Miller et al.~\cite{Miller:2019cac} and Riley et al.~\cite{Riley:2019yda}. However,\citet{Vinciguerra:2023qxq} conducted a re-analysis of the J0030 dataset using an upgraded pulse profile modeling pipeline and an improved instrument response model, incorporating background constraints. This re-analysis suggests that PSR J0030+0451 may exhibit more complex behavior than previously thought, with different modes corresponding to distinct hot spot geometries\footnote{The hot spots, which give rise to the pulsation as the star rotates and the thermal emission from the magnetic poles of the star, are thought to arise due to the heat generated from magnetospheric return currents \citep[see, e.g.,][]{1975ApJ...196...51R,NANOGrav:2017wvv,Harding:2001at,Salmi:2020iwq}.}, leading to varying inferred masses and radii. They adopted four different models to describe increasing complexity in the hot spot shapes: \texttt{ST-U}, \texttt{ST+PST}, \texttt{ST+PDT}, and \texttt{PDT-U}. Among these, both the \texttt{ST+PDT} and \texttt{PDT-U} models are capable of reproducing both \textit{NICER} and \textit{XMM-Newton} data, with the latter being significantly preferred by Bayesian evidence. The \texttt{PDT-U} model yielded a radius of \( R = 14.44^{+0.88}_{-1.05} \, \mathrm{km} \) and a mass of \( M = 1.70^{+0.18}_{-0.19} \, M_{\odot} \). The \texttt{ST+PDT} model, on the other hand, produced mass and radius estimates of \( M = 1.40^{+0.13}_{-0.12} \, M_{\odot} \) and \( R = 11.71^{+0.88}_{-0.83} \, \mathrm{km} \). These values are consistent with the magnetic field geometry inferred for the gamma-ray emission of this source~\citep{Kalapotharakos:2020rmz} and align closely with the results for PSR J0437+4715. Throughout this paper, we primarily present our results based on the \texttt{ST+PDT} model, as its inferred properties are consistent with independent gamma-ray observations and provide a good match with other well-studied pulsars. However, in Section~\ref{subsection:J0030_systematics}, we will compare these results with those obtained using the \texttt{PDT-U} model to assess the impact of these systematics on our analysis.

For PSR J0740+6620, the availability of a larger NICER data set has enabled more robust constraints on the mass and radius. The latest joint analysis of NICER and XMM data by \citet{Salmi:2024aum} reports a mass of $M = 2.07 \pm 0.07\, M_{\odot}$ and a radius of $R = 12.49_{-0.88}^{+1.28}$ km. 

The recent NICER observations of PSR J0437+4715~\cite{Choudhury:2024xbk} further refine our understanding of the EOS by providing additional constraints. This pulsar’s mass has been measured at $1.418 \pm 0.037 \,M_{\odot}$, with an equatorial radius of $11.36^{+0.95}_{-0.63}$ km. These results, derived through a combination of NICER and PPTA data, suggest a softer EOS~\cite{Rutherford:2024srk} and contribute new data that enhance the overall constraints on the properties of dense matter in neutron stars.

For each observation made by NICER, we construct the likelihood in the following manner,
\begin{align}
    P(D_{\rm X-ray}|\theta) = \int^{M_{\mathrm{max} }}_{M_{\mathrm{min} }} dM P(M|\theta) \nonumber \\ \times
    P(D_{\rm X-ray} | M, R (M, \theta)) \,.
\end{align}

\subsection{pQCD Likelihood}
\label{subsec:pqcd_input}

Due to asymptotic freedom, Quantum Chromodynamics (QCD) can be treated perturbatively at high densities beyond approximately $40$ times nuclear saturation density \cite{Andersen:2002jz}. Recent discussions have highlighted the nontrivial constraints that pQCD can impose on the NS EOS when advanced N3LO perturbative results are extrapolated to NS densities, incorporating stability, causality, and consistency arguments \cite{Gorda:2022jvk,Komoltsev:2021jzg}. The formalism was initially introduced in Ref.~\cite{Komoltsev:2021jzg}, and we provide a concise overview here, directing readers to the original work for in-depth details.

Consider an EOS characterized by a correlated set of values $\vec{\beta} \equiv \{p(\mu),n(\mu),\mu\}$, where $p$ is the pressure, $n$ is the number density, and $\mu$ is the chemical potential. Suppose knowledge of the EOS is available at limiting values of low density, $\mu_{\text{low}}$, and high density, $\mu_{\text{high}}$, such that
\begin{align}
   \vec{\beta}_{\text{low}} &= \{p_{\text{low}},n_{\text{low}},\mu_{\text{low}}\}  \equiv \{p(\mu_{\text{low}}),n(\mu_{\text{low}}),\mu_{\text{low}}\}, \label{eq:betalow}\\
   \vec{\beta}_{\text{high}} &= \{p_{\text{high}},n_{\text{high}},\mu_{\text{high}}\} \equiv \{p(\mu_{\text{high}}),n(\mu_{\text{high}}),\mu_{\text{high}}\}. \label{eq:betahigh}
\end{align}

While there are infinite EOSs connecting $\vec{\beta}_{\text{low}}$ and $\vec{\beta}_{\text{high}}$, any such EOS must adhere to thermodynamic stability, causality, and consistency. Thermodynamic stability requires concavity of the grand-canonical potential ($\Omega$) with respect to $\mu$, implying $\partial_\mu^2 \Omega \leq 0$~\cite{3097}. At $T=0$, $\Omega(\mu) = -p(\mu)$, and $n = \frac{\partial p}{\partial \mu}$, resulting in a stability constraint on the slope of $n(\mu)$: $\frac{\partial n}{\partial \mu} \geq 0$. Here, the derivative \(\frac{\partial}{\partial \mu}\) is taken at fixed temperature \(T\)  (which is \(T = 0\) in this case). Causality demands $c_s^2 \leq 1$, and at $T=0$, this relates to $n(\mu)$ and $\partial_\mu n$ through $c_s^{-2} = \frac{\mu}{n}\frac{\partial n }{\partial \mu } \geq 1$. Combining stability and causality, the slope of the curve for a maximally stiff EOS ($c_s^2 = 1$) at each point in $\mu - n$ space is $\frac{\partial n}{\partial \mu} = \frac{n}{\mu}$.

Ensuring pressure conditions at $(\mu_{\text{low}},n_{\text{low}})$ and $(\mu_{\text{high}},n_{\text{high}})$, it must hold that
\begin{equation}\label{eq:integral_constraints}
    \int_{\mu_{\text{low}}}^{\mu_{\text{high}}} n(\mu)d\mu = p_{\text{high}} - p_{\text{low}} = \Delta p. 
\end{equation}

Constraints on $\Delta p$ based on stability and causality constraints on $n(\mu)$ can be obtained. A lower bound, $\Delta p_{\text{min}}$, is derived by identifying the curve that minimizes the integral in Eq.~\eqref{eq:integral_constraints} while respecting stability and causality. Similarly, an upper bound, $\Delta p_{\text{max}}$, corresponds to the curve maximizing the integral. Assuming $c_s^2$ is bounded from above only by the causal limit, we have \cite{Komoltsev:2021jzg}
\begin{equation}\label{eq:plow}
    \Delta p_{\text{min}}  = \frac{1}{2}\left(\frac{\mu_{\text{high}}^2}{\mu_{\text{low}}} - \mu_{\text{low}}\right)n_{\text{low}}
\end{equation}
\begin{equation}\label{eq:phigh}
    \Delta p_{\text{max}}  = \frac{1}{2}\left( \mu_{\text{high}} - \frac{\mu_{\text{low}}^2}{\mu_{\text{high}}}\right)n_{\text{high}}.
\end{equation}
All these constraints combined imply that for any $\vec{\beta}_{\text{high}}$ and for a fixed $(\mu_{\text{low}}, n_{\text{low}})$, $p_{\text{low}}$ must be between $[p_{\text{high}} - \Delta p_{\text{min}}, p_{\text{high}} - \Delta p_{\text{max}}]$.

These guidelines for connecting two arbitrary regimes via an EOS which respects stability, causality, and consistency can be used to extrapolate pQCD results to the lower densities relevant for NSs. That is because, if we know $\vec{\beta}_{\text{high}} = \vec{\beta}_{\text{pQCD}}$, we can check if an EOS for which we only have knowledge up to a lower matching density $n_{\text{low}} = n_{\text{match}}$ can be connected to $\vec{\beta}_{\text{pQCD}}$ through a causal and stable EOS.

Our knowledge from pQCD is derived from current state-of-the-art calculations in Refs.~\cite{Gorda:2018gpy,Gorda:2021gha}, which report a partial N3LO order calculation of the zero-temperature, high-density QCD grand-canonical potential. Because these results arise from a series expansion in the QCD coupling constant and are then truncated at a finite order, we have to estimate the truncation error introduced by the missing higher-order terms. In the case of QCD, the truncation error depends on a residual, unphysical renormalization scale, $\Bar{\Lambda} \propto \mu$, which is underdetermined. Instead, the standard approach is to vary $\Bar{\Lambda}$ around a fiducial scale by some fixed factor. We follow Ref.~\cite{Gorda:2022jvk}, which adopted a scale-averaging approach. That means that pQCD results are given as a family of independent predictions $\vec{\beta}_{\text{pQCD}}(X)$, where $X \equiv \frac{3\Bar{\Lambda}}{2\mu_{\text{high}}}$. We set $\mu_{\text{high}} = 2.6$ GeV based on Ref.~\cite{Fraga:2013qra}, which points out that the uncertainty estimation for pQCD calculations at this value is similar to that of $\chi$EFT at $1.1 n_{\text{sat}}$ (about $\pm 24\%$ variation around the mean value \cite{Gorda:2022jvk}). We consider $X \in [\frac{1}{2},2]$, the same range that was implemented in Ref.~\cite{Gorda:2022jvk} and that has been suggested by phenomenological models \cite{Schneider:2003uz,Rebhan:2003wn,Cassing:2007nb,Gardim:2009mt} as well as the large-flavor limit of QCD \cite{Ipp:2003jy}.

Now that we have defined the theoretical input from high densities, we need to discuss how we define the low-density input from our EOS model. For any NS EOS that we generate with hybrid nuclear+PP model, $k$, we have to check that it can be connected to $\vec{\beta}_{\text{pQCD}}(X)$, for a given $X$, from $\vec{\beta}_{\text{low}} = \{p_k(n_{\text{match}}), n_{\text{match}}, \mu_k(n_{\text{match}}) \}$. In practice, we check that $p_k(n_{\text{match}})$ leads to $\Delta p \ \in [\Delta p_{\text{min}}, \Delta p_{\text{max}}]$, given $p_{\text{high}}$ from $\vec{\beta}_{\text{pQCD}}(X)$. Since the relevant scale for the NS EOS is the central density of a maximally massive star, $n_B^{\text{max}}$, we set $n_{\text{match}} = n_{B,k}^{\text{max}}$, which varies for each EOS. For the renormalization scale parameter, we follow Ref.~\cite{Gorda:2022jvk}, and sample $1,000$ values of $X \in [\frac{1}{2},2]$, evenly spaced in $\log(X)$. Hence, the pQCD weight associated with EOS $k$ is
\begin{equation}\label{eq:w_pQCD_def}
    w_{\text{pQCD}}(k) = \frac{1}{1000}\sum_{i= 1}^{1000} \mathbf{1}_{X}(k),
\end{equation}
where $\mathbf{1}_{X}(k)$ is the indicator function
\begin{equation}
    \mathbf{1}_{X}(k) = \begin{cases}
    1,& \text{if } \Delta p_k \in [\Delta p_{\text{min}}, \Delta p_{\text{max}}]\\
    0,              & \text{otherwise}
\end{cases},
\end{equation}
and $\Delta p_k = p_{\text{pQCD}}(X) - p_k(n_{B,k}^{\text{max}})$. Recall that $\Delta p_{\text{min}}$ and $\Delta p_{\text{max}}$ can be obtained from Eqs.~(\ref{eq:plow}, \ref{eq:phigh}), using 
\begin{align*}
    \mu_{\text{high}}  & =  2.6 \textrm{ GeV}, \\
    \mu_{\text{low}}  &=   \mu_k(n_{B,k}^{\text{max}}),\\
    n_{\text{high}}  &=   n_{\text{pQCD}}(\mu = 2.6 \textrm{ GeV},X), \\
    n_{\text{low}}  &=  n_{B,k}^{\text{max}}.
\end{align*}
Effectively, $w_{\text{pQCD}}(k)$ captures how often, out of the $1,000$ values for $X$, EOS $k$ can be connected to $\vec{\beta}_{\text{pQCD}}(X)$ with an EOS that respects thermodynamic stability, causality, and consistency. This procedure defines a weighting scheme associated with input from pQCD, which suppresses EOSs that are in tension with pQCD results by a factor proportional to the strength of the disagreement under the scale-averaging assumption.


As for the implementation of the likelihood, we use the public code released on \href{https://github.com/OKomoltsev/QCD-likelihood-function}{Github} \citep{Gorda:2021znl,Gorda:2022jvk,Komoltsev:2021jzg}.

\subsection{Neutron skin measurement likelihood} 
\label{subsection: skin_likelihood}
PREX-II collaboration has reported~\cite{PREX:2021umo} the value of neutron skin thickness of $^{208} {\rm Pb}$ to be, $R_{\rm skin}^{208} = 0.29 \pm 0.07$ fm (mean and $1 \sigma$ standard deviation). Ref.~\cite{Reed:2021nqk} have argued, this result implies the value of $L$ to be $106 \pm 37$ MeV. They also deduce a larger value of the NS radius as there exists a correlation between $R_{1.4}$ and $L$ based on relativistic mean field  calculations. However, Ref.~\cite{Biswas:2021yge} argued that the measurement uncertainty in $R_{\rm skin}^{208}$ by PREX-II is rather broad. Moreover, the measurement uncertainty deduced after the addition of astrophysical observations is dominated by the latter.
Combined astrophysical observations and PREX-II data yield the value of empirical parameter $L = 69^{+21}_{-19}$ MeV, $R_{\rm skin}^{208} = 0.20_{-0.05}^{+0.05}$ fm, and radius of a $1.4 M_{\odot}$ ($R_{1.4}) = 12.70_{-0.54}^{+ 0.42}$ km  at $1 \sigma$ credible interval (CI). Nevertheless, a better measurement of $R_{\rm skin}^{208}$ might have a small effect on the radius of low mass NSs, but for the high masses, there will be almost no effect. State-of-the-art $\chi$EFT calculations predict~\cite{Xu:2020fdc} the value of $R_{\rm skin}$ to be $0.17-0.18$ fm based on Bayesian analyses using mocked data. Therefore, if the high $R_{\rm skin}^{208}$ values continue to persist with lesser uncertainty, it would pose a challenge to the current theoretical understanding of nuclear matter near  saturation density. More recently, the CREX collaboration~\cite{CREX:2022kgg} has reported the value of neutron skin thickness of $\rm ^{48} \! Ca$ to be $R_{\rm skin}^{48} = 0.121 \pm 0.026$ (exp) $\pm 0.024$ (model) fm. They find  several models, including the microscopic
coupled cluster calculations~\cite{Hu:2021trw} are consistent with the combined CREX and PREX-II results at $90 \%$ CI, but in tension at $68 \%$ CI.
Following Ref.~\cite{Biswas:2021yge} we estimate the skin thickness $R_{\text{skin}}$ using the universal relation from Vi\~nas \textit{et al}.~\cite{Vinas:2013hua} connecting $R_{\text{skin}}$ and the empirical parameter $L$: $R_{\text{skin}}^{208} [\text{fm}] = 0.101 +  0.00147 \times L [\text{MeV}]$. 
Then we also use another empirical relation from Ref.~\cite{Tripathy:2020yig}, $R_{\text{skin}}^{48} =  0.0416 +  0.6169 R_{\text{skin}}^{208}$, to obtain the neutron skin thickness of $^{48} \text{Ca}$. The
likelihood for a given sample point is simply obtained
by comparing to the experimentally determined neutron skin thickness with the result predicted from the phenomenological relations set up by our model collection,

\begin{equation}
    P\left( D \mid \text{EOS} \right) \propto \exp\left( - \sum_i \frac{\left( \mathcal{R}_{\rm i}^{\rm model} - \mathcal{R}_{\rm i}^{\rm exp} \right)^2}{2 \sigma_i^2} \right),
\end{equation}

\section{Gaussian Mixture Model for Multi-Dimensional Observational Data}
\label{sec:GMM}
Gravitational wave detections from binary neutron star mergers offer a four-dimensional (4D) dataset, including the masses and tidal deformabilities of the two neutron stars involved. Concurrently, X-ray observations from the NICER collaboration yield a two-dimensional (2D) dataset, consisting of mass and radius measurements of neutron stars. Additionally, joint distributions of empirical EOS parameters such as \( L \) and \( K_{\rm sym} \), obtained from \(\chi\)EFT calculations, add another layer of complexity. To robustly model these complex and multi-dimensional datasets, we employ Gaussian Mixture Models (GMMs) using the \texttt{scikit-learn}~\cite{Planck:2015fie} library.

We fit Gaussian Mixture Models (GMMs) to various datasets and assess model performance using the Bayesian Information Criterion (BIC)~\cite{Kass:1995loi}. For the 4D dataset from gravitational wave detections, including parameters $M_1$, $M_2$, $\Lambda_1$, and $\Lambda_2$, and the 2D dataset from NICER X-ray observations, which includes neutron star mass and radius, we fit GMMs with varying numbers of components. The optimal model for each dataset is selected based on the BIC criterion, balancing model complexity with goodness of fit. Additionally, we model the joint distribution of empirical EOS parameters $L$ and $K_{\rm sym}$ derived from $\chi$EFT calculations using GMMs. The optimal number of components for this joint distribution is also chosen based on the BIC, capturing the complex dependencies between $L$ and $K_{\rm sym}$.

By optimizing the number of components, we ensure that the models are neither underfitted nor overfitted. The selected GMMs are used to compute the probability density of any given set of parameters.

\begin{figure*}[ht!]
    \centering
    \includegraphics[width=\textwidth]{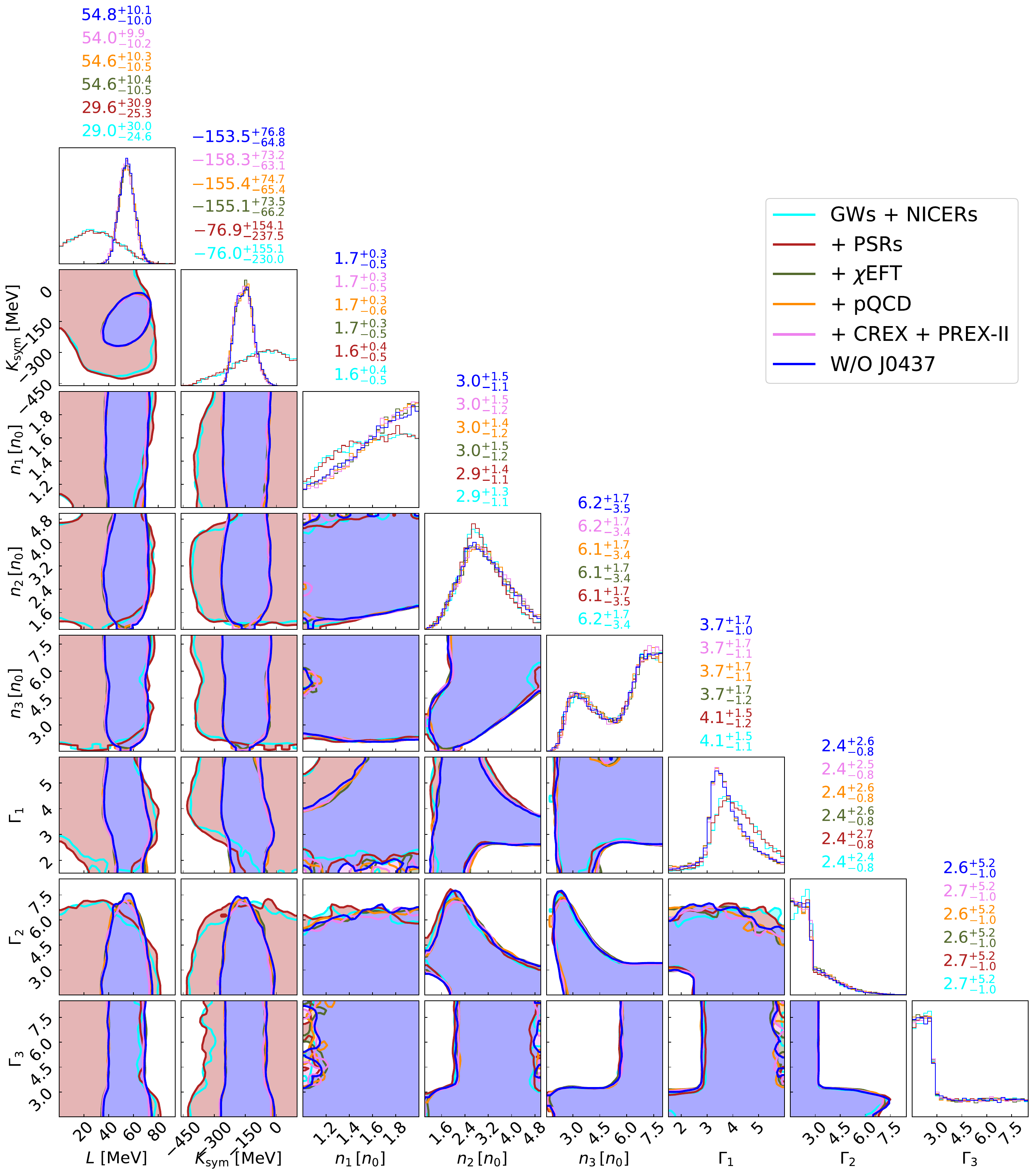}
    \caption{The posterior distributions of the EOS parameters are illustrated by sequentially adding different types of constraints, as indicated in the legend of the plot. In the marginalized one-dimensional plots, the median and 90\% CIs are shown for each constraint type, with each constraint represented in its respective color.}
    \label{fig:eos_params}
\end{figure*}

\begin{figure}[ht!]
    \centering

    \includegraphics[width=0.45\textwidth]{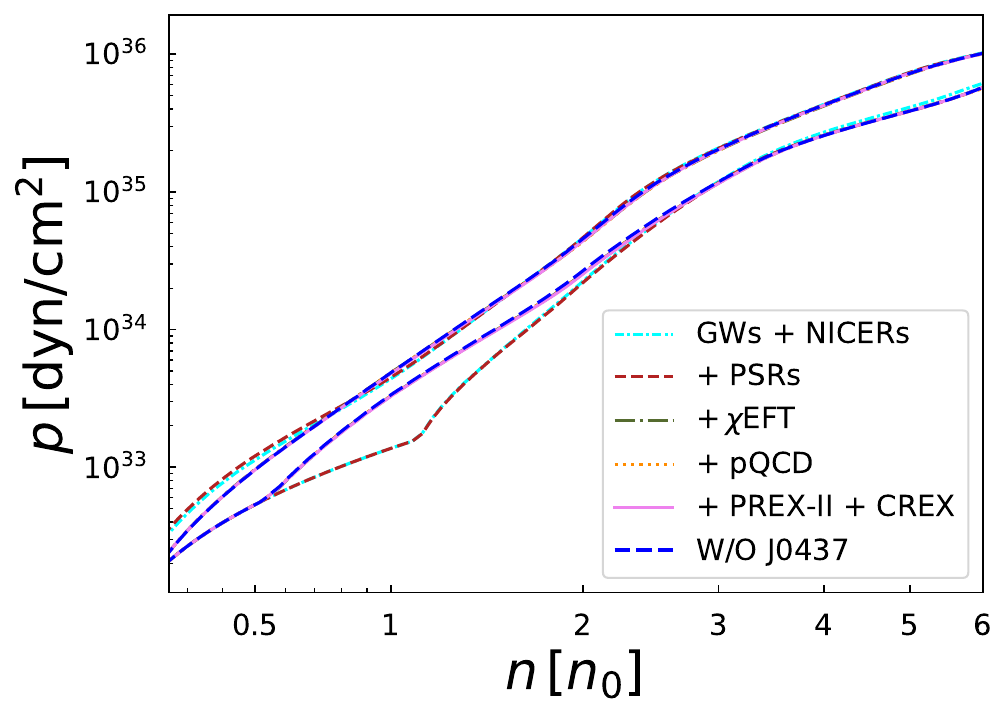} 
\caption{90 \% CI of the marginalized posterior distribution of the pressure in NS interior as a function of baryon density is shown by adding different types of constraints successively one after another, as indicated in the legend of the plot.}
    \label{fig:eos-post}
\end{figure}

\begin{figure}[ht!]
    \centering
    \includegraphics[width=0.45\textwidth]{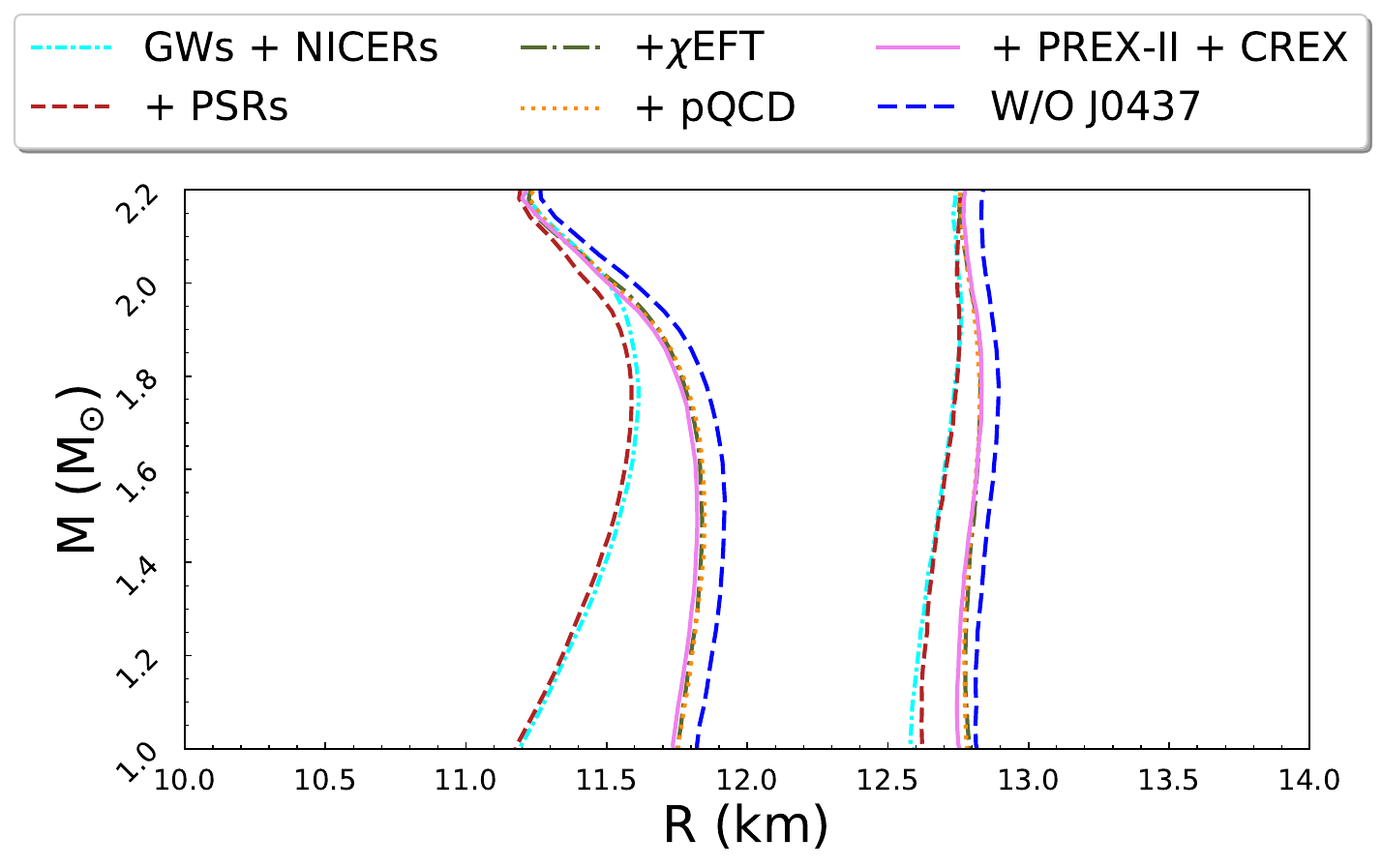}
  
\caption{90 \% CI of the marginalized posterior distribution of mass vs radius is shown by adding different types of constraints successively one after another, as indicated in the legend of the plot.}
    \label{fig:mr-post}
\end{figure}

\begin{figure*}[ht!]
    \centering
    \includegraphics[width=\textwidth]{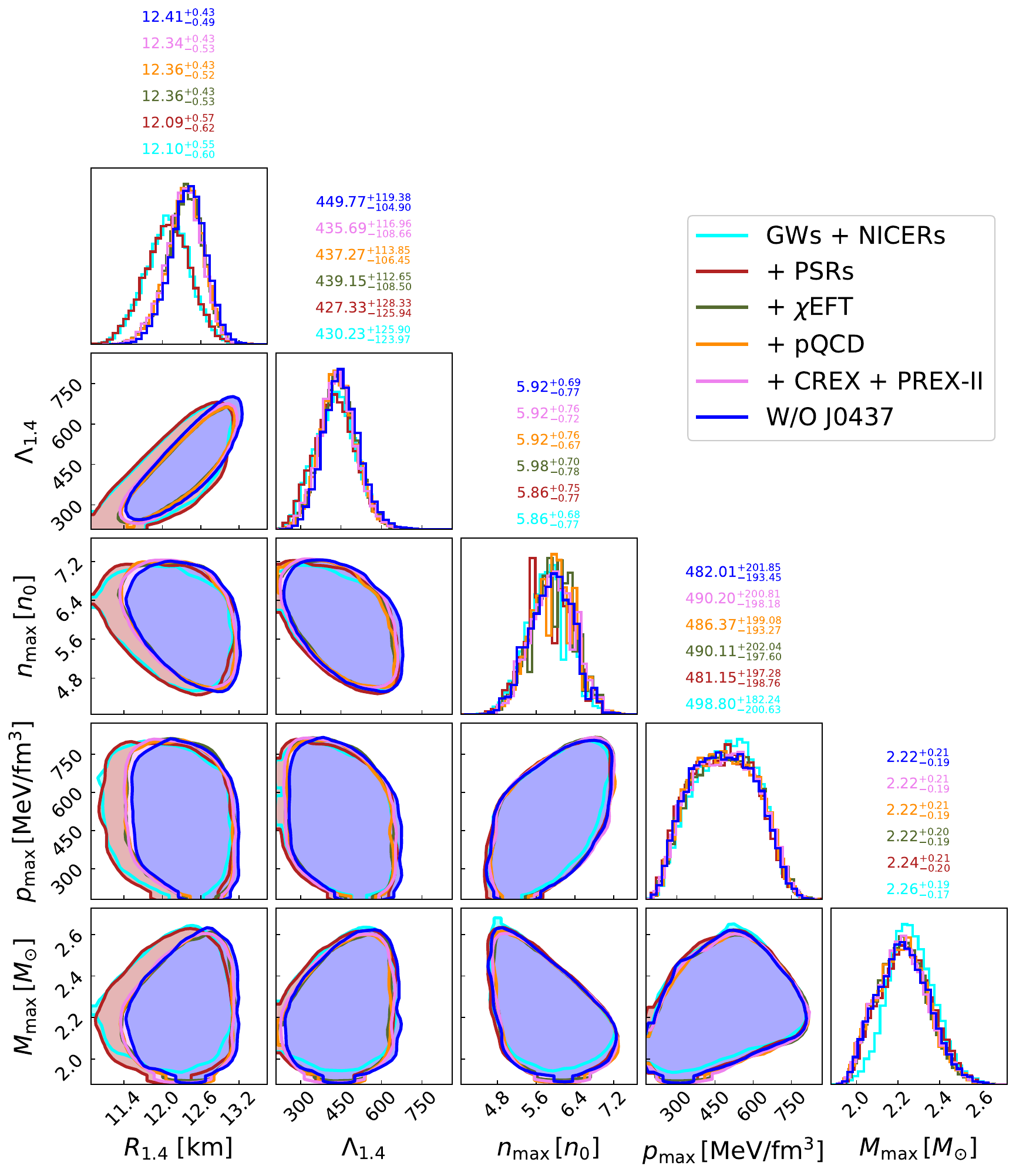}
\caption{Correlations between radius ($R_{1.4}$) and tidal deformability ($\Lambda_{1.4}$) of a $1.4M_{\odot}$ NS, maximum density ($n_{\rm max}$), maximum pressure ($p_{\rm max}$), and maximum mass ($M_{\rm max}$) are shown by adding different types of constraints successively one after another, as indicated in the legend of the plot. In the marginalized one-dimensional plots, the median and 90\% CIs are shown for each constraint type, with each constraint represented in its respective color.}
    \label{fig:macro_params}
\end{figure*}

\begin{figure}[ht!]
    \centering
    \includegraphics[width=0.45\textwidth]{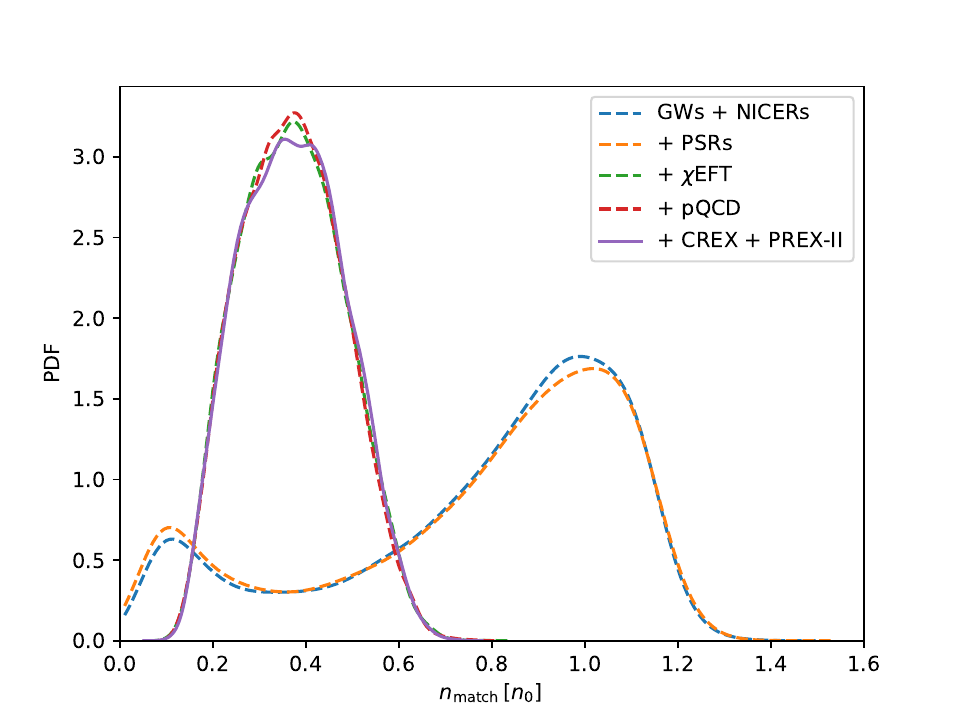}
\caption{Evolution of the inferred distribution of crust-core junction density as we add the constraints successively.}
    \label{fig:n_match}
\end{figure}

\begin{figure*}[ht!]
    \centering
    \includegraphics[width=\textwidth]{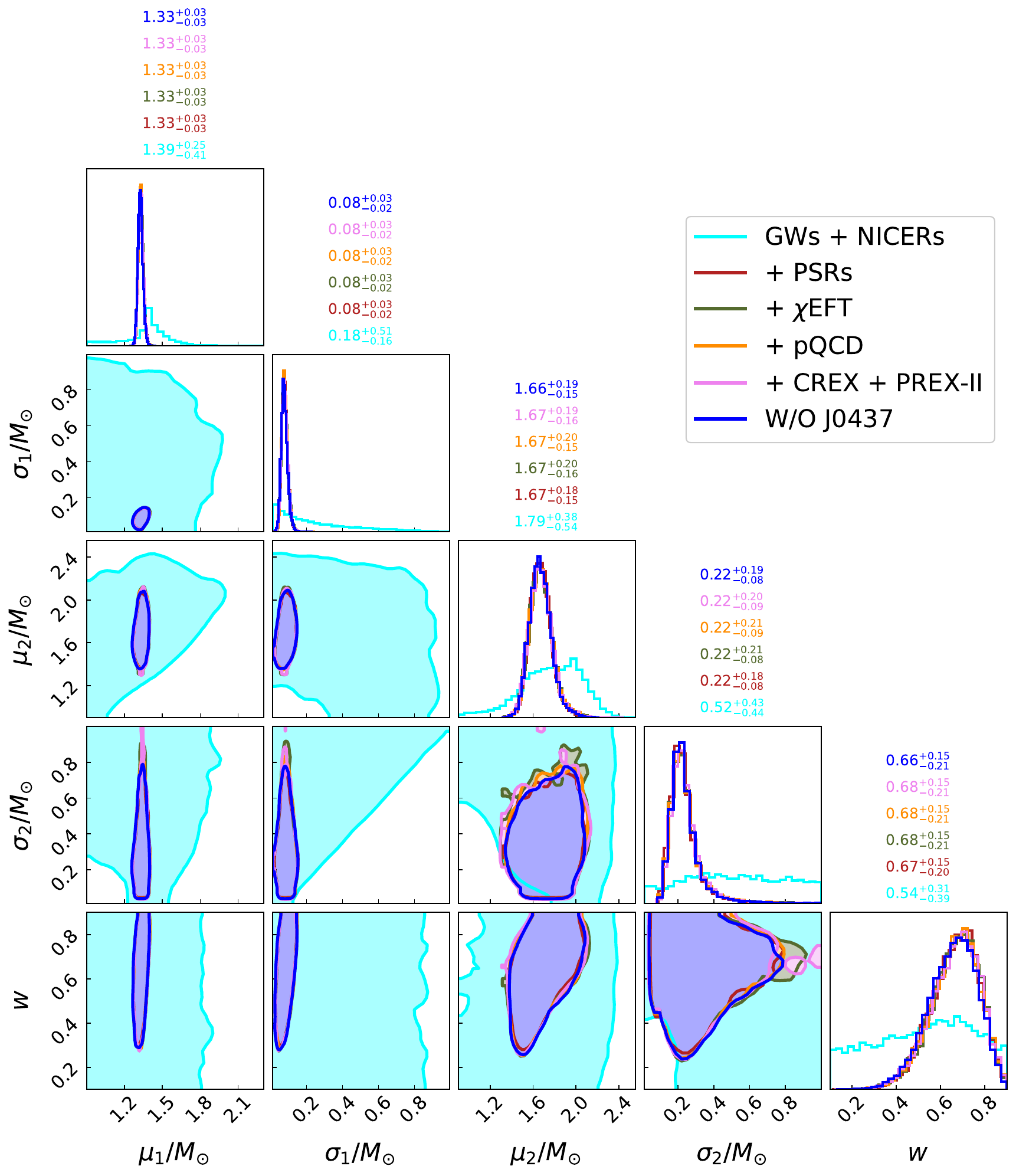}
\caption{Posterior distribution of mass model parameters are shown by adding different types of constraints successively one after another, as indicated in the legend of the plot. In the marginalized one-dimensional plots, the median and 90\% CIs are shown for each constraint type, with each constraint represented in its respective color.}
    \label{fig:pop_params}
\end{figure*}

\begin{figure}[ht!]
    \centering
    \includegraphics[width=0.45\textwidth]{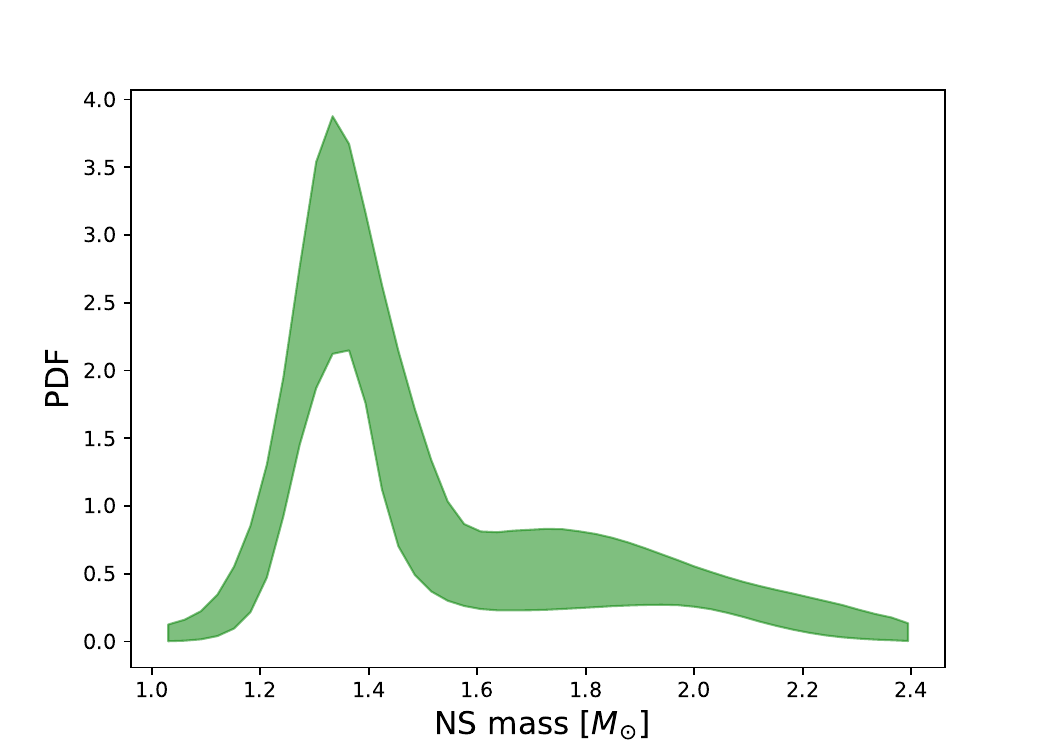}
\caption{Posterior distribution of total NS masses after adding the constraints coming from astrophysical observations, nuclear theory and experiments. }
    \label{fig:NS_mass_dist}
\end{figure}

\section{Results}
\label{section: results}
In Fig.~\ref{fig:eos_params}, the posterior distribution of all the EOS parameters is shown using the prior ranges from Table~\ref{tab:prior} by adding different types of constraints successively one after another.  It should be emphasized that we infer the EOS and the population parameters of NS simultaneously. However, for illustrative purposes, we do not show the joint posteriors of all the parameters, as they may not be adequately presented in the paper. In the marginalized one-dimensional plots, the median and  $90 \%$ CI are shown. In the legend of each corner plot the profiles of datasets are indicated. Initially, we combine the \(M-\Lambda\) measurements from GW170817 and GW190425 with the \(M-R\) measurements from the observations of PSR J0030+0451, PSR J0740+6620, and PSR J0437+4715. Following this, we incorporate the 70 reliable mass measurements of NSs as discussed in Ref.~\ref{subsection: mass-like}. Collectively, these gravitational wave, NICER, and pulsar (PSR) observations  represent joint constraints derived from astrophysical observations. Thereafter, we add the constraints coming from the theoretical calculations predicted by the $\chi$EFT prescription along with all astrophysical observations. This is shown using green colour. In the next step, constraints coming from pQCD calculations are added with astrophysical observations and $\chi$EFT prediction, which is shown in orange colour. Finally, we add the constraints coming from CREX and PREX-II experiments shown in pink colour. Since the mass and radius measurements of PSR J0437+4715 are very recent, we discuss their impact separately. The blue color represents the combination of all astrophysical observations and nuclear physics constraints, excluding PSR J0437+4715, and is labeled as `W/O J0437'. The resulting marginalized posterior distributions of pressure inside of NSs are shown in Fig.~\ref{fig:eos-post} as a function of energy density by adding successive observations, and in Fig.~\ref{fig:mr-post} corresponding mass-radius posterior distributions are shown. To obtain this plot, the 90\% CI of pressure (radius) is computed at a fixed energy density (mass) and those are plotted as a function of energy density (mass). Fig.~\ref{fig:macro_params} shows correlations between selected microscopic and macroscopic NS properties for analyses using different combinations of astrophysical and nuclear constraints. \\
The key points that we learn from these figures are the following:

\begin{itemize}
    \item \textbf{Astrophysical Observations Constrain Nuclear Parameters:} Astrophysical observations alone can place significant constraints on the nuclear empirical parameters, specifically the slope of the symmetry energy \(L\) and the curvature parameter \(K_{\rm sym}\). Based on the data, the median value of \(L\) is approximately 30 MeV with a 90\% CI of [5, 61] MeV. This shows that astrophysical observations provide a relatively tight constraint on \(L\), indicating its crucial role in defining the symmetry energy at nuclear saturation density. On the other hand, the median value of \(K_{\rm sym}\) is approximately \(-76\) MeV, with a broader CI of \([-306, 79]\) MeV. This broader range for \(K_{\rm sym}\), as expected from its nature as a quadratic term in the expansion, suggests that while the astrophysical data provide some information, the constraint on \(K_{\rm sym}\) is not as tight as that on \(L\). Notably, the data indicate a preference for a negative value of \(K_{\rm sym}\), which may suggest a softer symmetry energy at higher densities.

    \item \textbf{Significant Impact of \(\chi\)EFT Calculations on Empirical Parameters:} The inclusion of \(\chi\)EFT calculations has a profound impact on the constraints of the empirical parameters. After incorporating \(\chi\)EFT constraints, the median value of \(L\) becomes 55 MeV with a much narrower CI of [45, 65] MeV. This tighter bound on \(L\) highlights the power of \(\chi\)EFT in refining the density dependence of the symmetry energy, particularly in the low-density regime where \(\chi\)EFT is most applicable. For \(K_{\rm sym}\), the median value becomes to \(-155\) MeV, with a significantly reduced CI of \([-221, -82]\) MeV. This marked narrowing of the CI for \(K_{\rm sym}\) underscores the significant role of \(\chi\)EFT calculations in constraining this parameter, further supporting the notion that \(K_{\rm sym}\) is negative, which implies a softer symmetry energy in the corresponding density region.

    \item \textbf{Significant Impact of \(\chi\)EFT Calculations on various macroscopic properties:} The inclusion of \(\chi\)EFT constraints further refines the EOS, particularly by tightening the uncertainties in the key macroscopic parameters of neutron stars. With the addition of \(\chi\)EFT, the median $R_{1.4}$ shifts slightly to 12.36 km, with a further reduced CI of [11.83, 12.79] km. The median $\Lambda_{1.4}$ decreases slightly to 439, with the CI tightening to [330, 552]. This improvement shows the importance of \(\chi\)EFT in constraining the stiffness of the EOS. 

    \item \textbf{Influence of PSR J0437+4715 Observations on EOS Constraints:} 
    In the analysis, the pink-colored distribution in the figures represents the scenario where all available astrophysical observations and nuclear physics constraints are included, which encompasses GW data, NICER observations, 70 reliable pulsar mass measurements, \(\chi\)EFT predictions, and the neutron skin thickness measurements from PREX-II and CREX. The blue-coloured distribution, labelled as `W/O J0437', shows the constraints derived from the combination of all these data except for the recent measurements of PSR J0437+4715. The inclusion of PSR J0437+4715 data slightly shifts the EOS constraints toward softer models. Specifically, when comparing the pink and blue distributions, the inclusion of PSR J0437+4715 generally results in a minor reduction in the inferred radius and tidal deformability for a $1.4 M_{\odot}$ neutron star. For instance, the median radius \(R_{1.4}\) decreases slightly, indicating a preference for a softer EOS, which corresponds to a lower pressure at a given density in neutron star matter. Similarly, the tidal deformability \(\Lambda_{1.4}\) also shows a subtle reduction, consistent with the inference of a softer EOS. 

    \item \textbf{Correlation Between Densities in EOS Inference:} We observe a noticeable impact on the posterior distribution of $\Gamma_1$ when we introduce constraints from $\chi$EFT, despite the fact that $\Gamma_1$ is associated with a higher density than where the $\chi$EFT calculations are applied. This suggests the densities in different regions are correlated due to the parametric nature of our EOS model, the imposition of causality condition on the speed of sound, and the monotonically increasing behaviour of pressure to maintain the mechanical stability.

    \item \textbf{Minimal Influence of pQCD Input on EOS Inference:} The findings presented here indicate that the influence of pQCD input on the inference of the NS EOS is negligible. Despite recent reports suggesting otherwise \cite{Gorda:2022jvk,Gorda:2022jvk,Altiparmak:2022bke}, we aim to investigate why this discrepancy exists. Let us revisit Section \ref{subsec:pqcd_input}, where we examined the assumptions related to integrating pQCD input into the analysis. We assume knowledge of the EOS at a low-density limit (refer to Eqs. (\ref{eq:betalow}, \ref{eq:betahigh})), with each limit determined by three values specifying the density ($n$), chemical potential ($\mu$), and pressure ($p$). While $\mu_{\rm low}$ and $p_{\rm low}$ are obtained from the EOS in the low-density regime, a choice must be made for $n_{\rm low}$. This choice is crucial because $n_{\rm low} = n_{\rm matching}$, indicating that pQCD results will extend down to $n_{\rm low}$. Although this selection is theoretically arbitrary, opting for $n_{\rm low} = n_B^{\rm max}$ for each EOS is rational, as $n_B^{\rm max}$ represents the highest relevant density scale for the EOS of isolated, slowly-rotating NSs. This is the approach adopted in our study and was also utilized in Ref.~\cite{Somasundaram:2022ztm,Mroczek:2023zxo}, which similarly concluded that pQCD exerted minimal influence on the prior EOS.

    \item \textbf{Neutron Skin Thickness Measurements and EOS Inference:} Similarly, we find that the neutron skin thickness measurements provided by the PREX-II and CREX collaborations have a minimal impact on the inference of the neutron star EOS. Although PREX-II alone suggests a relatively large value for the symmetry energy slope parameter \(L\) (as discussed in Section~\ref{subsection: skin_likelihood}), its influence diminishes when combined with other datasets. This suggests that the uncertainties in neutron skin thickness measurements from PREX-II and CREX are quite broad compared to the more precise constraints from other data, especially those from \(\chi\)EFT. As a result, the overall impact of PREX-II and CREX is currently overshadowed by the more dominant contributions from \(\chi\)EFT predictions and astrophysical observations. 

    \item \textbf{Crust-core junction density:} In our study, the crust EOS SLy is modeled using a piecewise polytropic representation, with the corresponding parameters taken from Read et al.~\cite{Read:2008iy}. The last polytropic index of SLy is applied at a density of \( 2.62789 \times 10^{12} \) g/cm\(^3\). The matching density, \( n_{\rm match} \), lies between this point and \( n_1 \), and it is determined by ensuring continuity in pressure between the SLy crust and our empirical EOS model. To further examine the impact of \( n_{\rm match} \), we have quantified its posterior distribution for each dataset, applying successive constraints. The results are presented in Fig.~\ref{fig:n_match}, where we show the evolution of the posterior as additional constraints are incorporated, including gravitational waves (GWs), NICER observations, pulsar mass measurements (PSRs), chiral effective field theory (EFT), perturbative QCD (pQCD), and nuclear experiments such as CREX and PREX-II. From the figure, it is evident that the inclusion of chiral EFT constraints has a significant impact on the inferred transition density, shifting the posterior towards lower values of \( n_{\rm match} \). This suggests that the crust-core transition is strongly influenced by nuclear physics constraints at sub-saturation densities.

\end{itemize}

In Fig.~\ref{fig:pop_params}, the posterior distribution of all the mass distribution model parameters are shown similarly as in Fig.~\ref{fig:eos_params}. The resulting distribution of observed NS masses is shown in Fig.~\ref{fig:NS_mass_dist}. Among the various datasets used, it is evident that the mass measurements of PSRs play the most significant role in determining these parameters. This is because PSR mass measurements offer direct, precise data on the masses of NSs, which strongly influence the inferred population characteristics.

Conversely, the impact of other constraints, such as those from GWs, NICER observations, $\chi$EFT, and nuclear physics constraints (including CREX and PREX-II), appears to be less influential on the posterior distribution of the mass population parameters. The posterior distributions remain relatively stable even after the inclusion of these additional constraints, which suggests that while these data provide valuable information for EOS inference, they do not significantly alter the mass population parameters derived primarily from PSR mass measurements.

The limited influence of these other constraints can be attributed to their indirect relationship with the mass distribution. For instance, GWs and NICER observations provide constraints on the radius and tidal deformability of NSs, which are related to the EOS but do not directly inform the mass distribution in the same way as precise PSR mass measurements. Similarly, nuclear constraints such as those from $\chi$EFT and neutron skin measurements (CREX and PREX-II) primarily influence the low-density behavior of the EOS and have broader uncertainties that do not substantially modify the mass population parameters.

The posterior distributions presented in those figures for the double Gaussian model parameters provide significant insights into the mass distribution of NSs. The model includes two Gaussian components characterized by their means (\(\mu_1\) and \(\mu_2\)), standard deviations (\(\sigma_1\) and \(\sigma_2\)), and a weight factor (\(w\)) representing the relative contribution of each Gaussian to the overall distribution.

\begin{itemize}
    \item \textbf{First Gaussian Component (\(\mu_1\) and \(\sigma_1\))}: The mean mass (\(\mu_1 = 1.33^{+0.03}_{-0.03} M_\odot\)) is tightly constrained, indicating a high degree of confidence that a significant portion of NSs cluster around this mass. The narrow spread (\(\sigma_1 = 0.08^{+0.03}_{-0.02} M_\odot\)) suggests that the mass distribution around this mean is relatively tight.
    
    \item \textbf{Second Gaussian Component (\(\mu_2\) and \(\sigma_2\))}: The mean mass (\(\mu_2 = 1.67^{+0.19}_{-0.16} M_\odot\)) is less tightly constrained compared to \(\mu_1\), reflecting greater uncertainty. The broader spread (\(\sigma_2 = 0.22^{+0.20}_{-0.09} M_\odot\)) indicates a wider range of masses around this mean, suggesting more diversity in the mass values in this component.
    
    \item \textbf{Weight Factor (\(w\))}: The weight \(w = 0.68^{+0.15}_{-0.21}\) indicates that the first Gaussian component contributes slightly more than half of the total NS population. This balance highlights the presence of two distinct populations or formation channels for NSs.
\end{itemize}

The tight constraint on \(\mu_1\) and the narrow spread \(\sigma_1\) support the hypothesis that a large fraction of NSs are formed with masses around 1.33 \(M_\odot\), which aligns with observations of NSs in binary systems~\cite{Kiziltan:2013oja}. The broader distribution for the second component suggests a more heterogeneous formation process, potentially involving a wider variety of progenitor masses or post-formation mass changes such as accretion in binary systems~\cite{10.1111/j.1365-2966.2009.14963.x,Vigna-Gomez:2018dza}.

\section{Systematic Uncertainties in EOS Inference}
\label{sec:systematics}
In the context of NS EOS inference, several sources of systematic uncertainties must be considered, particularly those arising from the assumptions made in modeling the EOS and the mass population of neutron stars. While the statistical uncertainties are well quantified, the potential systematic biases could influence the final constraints on the EOS parameters.

\subsection{Systematics in Mass-Radius Posterior Distributions of PSR J0030+0451}
\label{subsection:J0030_systematics}
The reanalysis of NICER data for PSR J0030+4510 conducted by Ref.~\cite{Vinciguerra:2023qxq} demonstrates a strong dependence on the choice of hotspot emission model. As discussed in Sec.~\ref{subsec: NICER_Likelihood}, both the \texttt{ST+PDT} and \texttt{PDT-U} models are capable of reproducing \textit{NICER} and \textit{XMM-Newton} data. These models yield notably different mass and radius estimates for PSR J0030+0451, which, in turn, influence the inferred NS equation of state.

\begin{figure}
    \centering
    \includegraphics[width=0.45\textwidth]{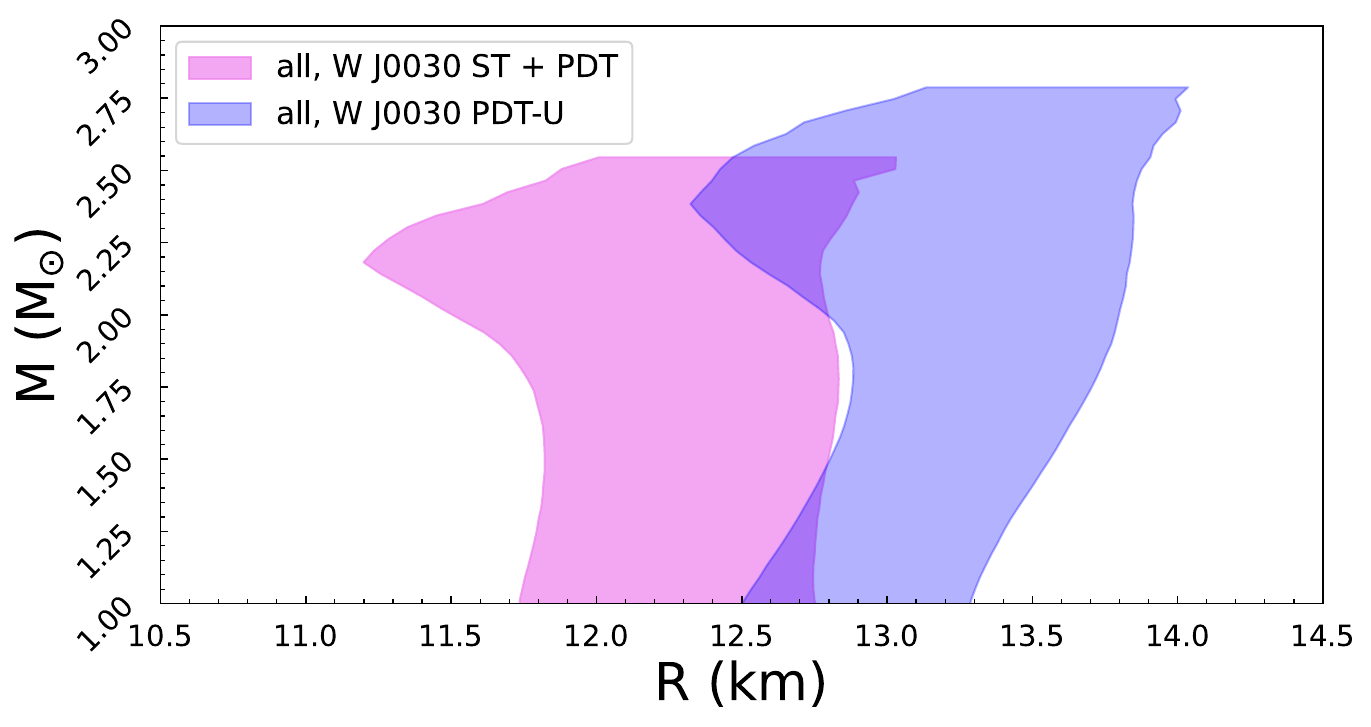}
    \caption{Mass-radius posterior distributions for neutron stars obtained using the ST+PDT and PDT-U models, combining astrophysical and nuclear constraints. The shaded regions represent the 90\% credible intervals.}
    \label{fig:mr_J0030_systematics}
\end{figure}

The ST+PDT model provides a relatively softer posterior distribution in the mass-radius space. In contrast, the PDT-U model produces a significantly stiffer posterior, indicating larger radii for a given mass. The resultant 90\% CI mass-radius posterior distributions, combining other astrophysical and nuclear constraints, are shown in Figure~\ref{fig:mr_J0030_systematics}. 
A key observation is the limited overlap between the posterior distributions obtained from the two models. This small overlap highlights the substantial influence of model-dependent systematics in constraining the NS EOS. The stiffness of the PDT-U posterior suggests a preference for equations of state that are more resistant to compression, leading to larger NS radii. Conversely, the softer ST+PDT posterior aligns with EOS predictions favoring more compact neutron stars.

These differences illustrate the necessity of careful consideration when selecting models for NICER data analysis. The model-dependent variations emphasize the need to incorporate systematic uncertainties into Bayesian inference pipelines to ensure robust conclusions about the properties of dense matter.

The stark differences between these two modeling approaches also have implications for multimessenger studies of neutron stars. For instance, tighter constraints on the EOS derived from PDT-U might be less consistent with those inferred from gravitational wave data, which often favor softer equations of state. Future analyses will benefit from combining these models with additional multimessenger observations to refine mass-radius constraints and mitigate model-dependent biases.

~\subsection{Impact of Including Less Reliable Pulsar Mass Measurements on EOS Constraints}
\label{subsection:PSR_systematics}
Throughout this paper, we have predominantly relied on highly reliable mass measurements of pulsars to constrain the neutron star equation of state. These measurements, when combined with other astrophysical observations and nuclear physics inputs, provide robust constraints and form the foundation for deriving credible mass-radius posterior distributions. However, in this section, we examine the effect of incorporating less reliable pulsar mass measurements to explore how these additional data points influence our results.

\begin{figure} \centering \includegraphics[width=\linewidth]{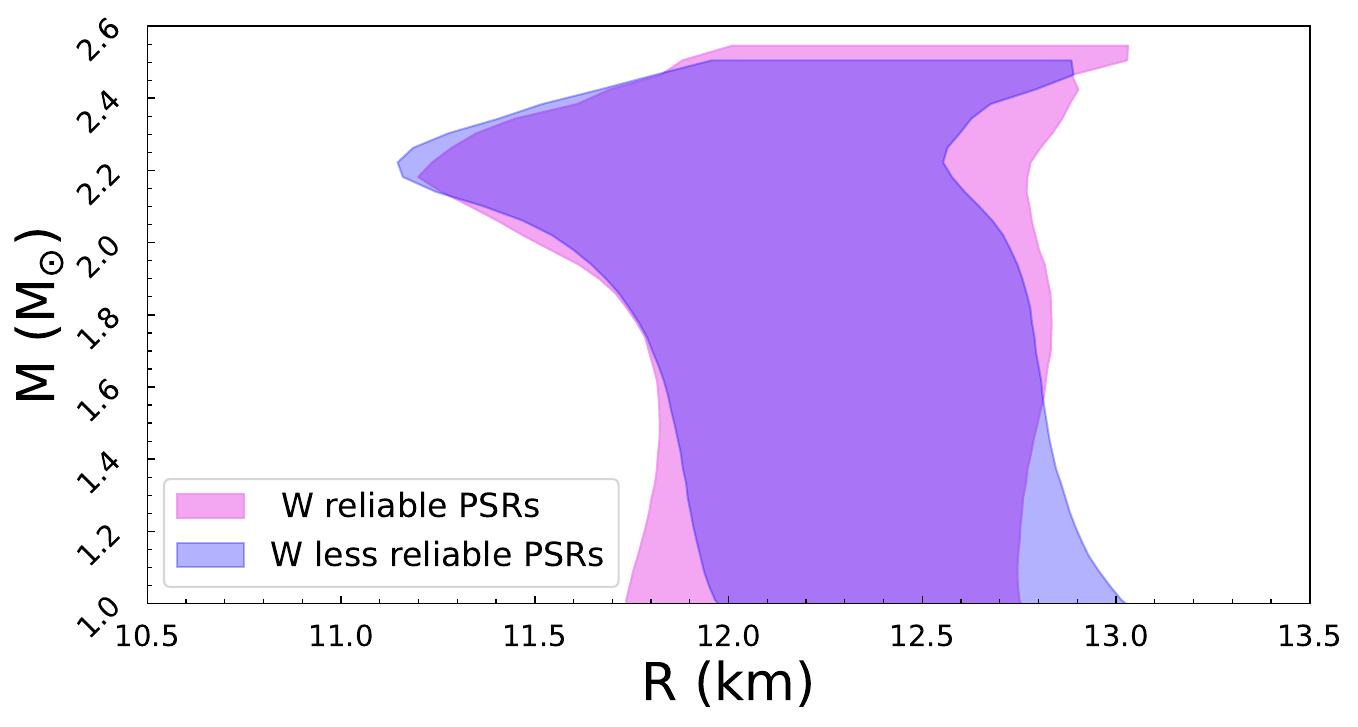} \caption{Comparison of mass-radius posterior distributions using only reliable pulsar mass measurements (orange) and including less reliable measurements (blue). Both cases include additional astrophysical observations and nuclear physics inputs. The shaded regions represent the 90\% credible intervals.} \label{fig:mr_PSR_systematics} 
\end{figure}

Figure~\ref{fig:mr_PSR_systematics} compares the 90\% credible regions of the mass-radius posterior distributions for two cases: (i) using only highly reliable pulsar mass measurements, combined with astrophysical and nuclear constraints, and (ii) including both reliable and less reliable mass measurements along with the same astrophysical and nuclear inputs. We observe that the inclusion of less reliable pulsar mass measurements leads to significantly tighter constraints on the EOS. Specifically, the inferred maximum mass of a neutron star increases slightly and is better constrained with the inclusion of less reliable data. The maximum mass is determined to be $2.25^{+0.13}_{-0.10} \, M_\odot$, a much narrower credible interval compared to the results obtained using only reliable mass measurements. This improvement highlights the potential of leveraging less reliable measurements, despite their inherent uncertainties, to refine EOS constraints. The tighter constraint is likely due to the additional data points spanning a broader range of masses and radii, which effectively reduce the parameter space for viable EOS models. This result demonstrates that, while reliability is critical for robust inference, carefully incorporating less reliable data can provide complementary insights and improve overall constraints.

It is important to note that systematic uncertainties associated with less reliable measurements must be carefully accounted for in future analyses. Integrating such uncertainties into Bayesian pipelines will ensure that conclusions drawn about the EOS remain robust and consistent with the underlying data quality.

\subsection{Assumptions in EOS Modeling}

The EOS at higher densities remains one of the most uncertain aspects of nuclear astrophysics. In this work, the EOS has been modeled using a piecewise polytropic representation, which is a common method in the literature. However, this is not the only approach available. Alternative methods include the speed of sound parameterization~\cite{Greif:2018njt}, spectral parameterization~\cite{Lindblom:2010bb}, or even nonparametric methods~\cite{Landry_2020PhRvD.101l3007L}, each of which has its own strengths and limitations.

Different EOS models can lead to slightly different constraints on neutron star properties such as radius, tidal deformability, and maximum mass. While these variations are generally within the statistical uncertainties, they underscore the importance of considering multiple EOS models to fully capture the range of possible behaviors at supranuclear densities. For instance, the piecewise polytropic approach allows for flexibility in modeling the EOS but may introduce biases depending on the choice of segment boundaries. On the other hand, spectral parameterization offers a smooth, global description of the EOS, but it may not capture sharp phase transitions as effectively as piecewise models.

Given the current uncertainties in the high-density behavior of neutron star matter, it is crucial to interpret EOS constraints with an understanding of the underlying assumptions in the modeling approach. Future work could benefit from a systematic comparison of different EOS models to better quantify the impact of these assumptions on the inferred neutron star properties.

\subsection{Assumptions in the Mass Population Model}

This work assumes a bimodal mass distribution for all neutron stars observed in the universe, motivated by the double Gaussian distribution observed in the galactic population of neutron stars. However, there is no \textit{a priori} reason to assume that this mass distribution would be the same in other galaxies or in the population of neutron stars detected through GWs. The galactic neutron star mass distribution is well-studied, but the neutron stars observed through GWs could follow a different mass distribution due to differing formation histories or environments. Although the current sample of NSs detected through GWs is relatively small, which limits the impact of a potentially incorrect population assumption on the EOS inference, it is important to recognize that this assumption may introduce a systematic bias as the number of GW-detected neutron stars increases in the future. Therefore, future studies should consider alternative mass distribution models to assess their impact on the EOS constraints.

\subsection{Selection Effects}

Selection effects represent another potential source of systematic uncertainty in the inference of the neutron star EOS and mass distribution. The mass measurements used in this analysis are predominantly from neutron stars in binary systems, which may not be representative of the entire neutron star population, particularly isolated neutron stars.

Selection effects arise from several observational biases. For example, Shapiro delay, which provides some of the most precise mass measurements, is easier to detect in systems with short orbital periods and high inclinations. Similarly, spectroscopic observations are more feasible in relatively compact systems. Furthermore, gravitational wave detectors are more sensitive to heavier binary systems, potentially skewing the inferred mass distribution towards higher masses.

These selection effects could influence the inferred mass distribution and, consequently, the EOS constraints. However, these potential biases were not explicitly explored in this work. As the field progresses and more diverse neutron star systems are observed, it will become increasingly important to account for these selection effects to avoid systematic biases in the EOS inference.

\section{Conclusion}
\label{section: conclusion}

In this study, we have undertaken a comprehensive examination of the NS EOS by integrating data from a variety of sources, including theoretical models, multimessenger astronomical observations, and experimental results. Our approach employs a hybrid EOS formulation that merges a parabolic expansion-based nuclear empirical parameterization near nuclear saturation density with a PP model at higher densities.

Our findings highlight the significant impact of $\chi$EFT in constraining the EOS, particularly at densities near or below nuclear saturation density. In contrast, data from pQCD and nuclear experiments such as PREX-II and CREX had a lesser impact on the constraints of the EOS.

By incorporating a total of 70 (129) reliable (including less reliable)  available NSs mass measurements up to April 2023, including NS mass and radius measurements from X-ray emissions and gravitational wave data from BNS mergers, we have derived more precise estimates for key EOS parameters. Notably, the slope and curvature of the symmetry energy, the radius, and tidal deformability of a 1.4 $M_{\odot}$ NS, and the maximum mass of a non-rotating NS have been determined with improved accuracy.

This synthesis of diverse datasets within a hierarchical Bayesian framework underscores the intricate relationship between nuclear physics and astrophysical observations. The enhanced constraints on the NS EOS contribute to a deeper understanding of the fundamental properties of dense nuclear matter and the extreme conditions within NSs. These insights pave the way for future research and observations, promising further revelations about the enigmatic nature of these celestial objects.

Our study demonstrates the power of combining theoretical advancements with cutting-edge observational and experimental data to unravel the complexities of the NS EOS. As observational technologies and theoretical models continue to evolve, we anticipate even more precise characterizations of NS properties, further bridging the gap between nuclear physics and astrophysics.

\section*{Acknowledgements}
 BB thanks i) Shao Peng for providing a Python script to model the asymmetric normal distribution used in this work, ii) Praveer Tiwari and Christian Drischler for useful discussion regarding the implementation of $\chi$EFT likelihood, and iii) Philippe Landry and Prasanta Char for reading the manuscript carefully and making useful suggestions. Calculations were performed on the facilities of the North-German Supercomputing Alliance (HLRN) and at the SUNRISE HPC facility supported by the Technical Division at the Department of Physics, Stockholm University. BB  acknowledges the support from the Knut and Alice Wallenberg Foundation 
under grant Dnr.~KAW~2019.0112 and the Deutsche 
Forschungsgemeinschaft (DFG, German Research Foundation) under 
Germany's Excellence Strategy – EXC~2121 ``Quantum Universe'' –
390833306. SR has been supported by the Swedish Research Council (VR) under grant number 2020-05044, by the research environment grant “Gravitational Radiation and Electromagnetic Astrophysical Transients” (GREAT) funded by the Swedish Research Council (VR) under Dnr 2016-06012, by the Knut and Alice Wallenberg Foundation under grant Dnr. KAW 2019.0112, by Deutsche Forschungsgemeinschaft (DFG, German Research Foundation) under Germany’s Excellence Strategy - EXC 2121 “Quantum Universe” - 390833306 and by the European Research Council (ERC) Advanced Grant INSPIRATION under the European Union’s Horizon 2020 research and innovation programme (Grant agreement No. 101053985). 

\bibliography{mybiblio}
\end{document}